\RequirePackage{xcolor}
\documentclass[journal,twoside,web]{ieeecolor}
\usepackage{generic}
\usepackage[T1]{fontenc}
\usepackage[utf8]{inputenc}
\usepackage{cite}
\usepackage{amsmath,amssymb,bm}
\usepackage{graphicx}
\usepackage{booktabs,array,diagbox,makecell}
\usepackage{xspace}
\definecolor{customblue}{rgb}{0.36, 0.55, 0.75}
\usepackage[colorlinks=true,linkcolor=customblue, urlcolor=magenta,citecolor=customblue, anchorcolor=customblue]{hyperref}
\usepackage{orcidlink}

\usepackage{etoolbox}
\makeatletter
\patchcmd{\@makecaption}{\setbox\@tempboxa\hbox}{\sbox\@tempboxa}{}{}
\patchcmd{\@makecaption}{\setbox\@tempboxa\hbox}{\sbox\@tempboxa}{}{}
\newcommand{\ps@jbhiopening}{\ps@headings\let\rightmark\leftmark}
\makeatother

\newcommand{\name}{ADformer\xspace}
\newcommand{\std}[1]{{\scriptsize\ensuremath{\pm#1}}}
\hypersetup{
  pdftitle={ADformer: A Multi-Granularity Spatial-Temporal Transformer for EEG-Based Alzheimer Detection},
  pdfauthor={Yihe Wang, Nadia Mammone, Darina Petrovsky, Alexandros T. Tzallas, Francesco C. Morabito, and Xiang Zhang},
  pdfsubject={IEEE Journal of Biomedical and Health Informatics}
}

\begin{document}
\bstctlcite{IEEEexample:BSTcontrol}
\title{ADformer: A Multi-Granularity Spatial-Temporal Transformer for EEG-Based Alzheimer Detection}
\author{Yihe Wang\,\orcidlink{0009-0000-5220-5452},
  Nadia Mammone\,\orcidlink{0000-0003-4962-3500},
  Darina Petrovsky\,\orcidlink{0000-0002-3109-1633},
  \\Alexandros T.~Tzallas\,\orcidlink{0000-0001-9043-1290},
  Francesco C.~Morabito\,\orcidlink{0000-0003-0734-9136},
  and Xiang Zhang\,\orcidlink{0000-0001-5097-2113}%
  \thanks{Yihe Wang and Xiang Zhang are with the Department of Computer Science, University of North Carolina -- Charlotte, Charlotte, North Carolina 28223, United States. Corresponding author: Xiang Zhang (e-mail: \href{mailto:xiang.zhang@charlotte.edu}{xiang.zhang@charlotte.edu}).}%
  \thanks{Nadia Mammone and Francesco C. Morabito are with the DICEAM Department, University Mediterranea of Reggio Calabria, Via Zehender, Feo di Vito, 89122 Reggio Calabria, Italy.}%
  \thanks{Darina Petrovsky is with the School of Nursing, Duke University, Durham, North Carolina 27708, United States.}%
  \thanks{Alexandros T. Tzallas is with the Department of Informatics \& Telecommunications, University of Ioannina, Kostakioi, GR-47100, Arta, Greece.}%
  \thanks{Code: \url{https://github.com/DL4mHealth/ADformer}.}\\}
\markboth{IEEE Journal of Biomedical and Health Informatics}
  {Wang \MakeLowercase{\textit{et al.}}: ADformer for EEG-Based Alzheimer Detection}
\maketitle
\thispagestyle{jbhiopening}

\begin{abstract}
Electroencephalography has emerged as a cost-effective, efficient tool to help neurologists detect Alzheimer’s Disease. However, most existing approaches rely heavily on manual feature engineering or data transformation. While these techniques may help with small-scale datasets, they often cause information loss and distortion when applied to large-scale data, ultimately limiting model performance. Moreover, the limited subject scale and demographic diversity of datasets used in prior studies hinder comprehensive evaluation of model robustness and generalizability, thus restricting their applicability in real-world clinical settings. To address these challenges, we propose \textit{\name}, a novel multi-granularity spatial-temporal transformer that captures both temporal and spatial features from raw EEG signals, enabling effective end-to-end representation learning. Our model introduces multi-granularity embedding strategies across both spatial and temporal dimensions, leveraging a two-stage intra-inter-granularity self-attention mechanism to learn local patterns within each granularity and global dependencies across granularities. We evaluate \textit{\name} on 4 large-scale datasets totaling 1,713 subjects, one of the largest corpora for EEG-based AD detection to date, under a cross-validated, subject-independent setting. Experimental results demonstrate that \textit{\name} consistently outperforms existing methods, achieving subject-level F1 scores of 87.11\%, 89.83\%, 67.99\%, and 84.42\% on the 4 datasets, respectively, in distinguishing Alzheimer’s Disease patients from healthy control subjects.

\end{abstract}

\section{Introduction}
\label{sec:intro}

Alzheimer’s disease is the most prevalent neurodegenerative disorder among the elderly population worldwide~\cite{breijyeh2020comprehensive}, affecting an estimated 10-30\% of individuals over the age of 65, with an annual incidence rate of 1-3\%~\cite{masters2015alzheimer}. AD is characterized by the accumulation of amyloid-$\beta$ peptides in the brain due to impaired clearance mechanisms~\cite{murphy2010alzheimer}. Clinically, the disease progresses in stages, typically advancing from healthy aging to mild cognitive impairment (MCI) and ultimately to overt AD. Although a definitive cure for AD remains elusive, early detection and intervention can temporarily slow symptom progression and improve patients’ quality of life~\cite{nelson2015slowing, chu2012alzheimer}. Current AD diagnostic procedures typically combine clinical and cognitive assessments with neuroimaging techniques, such as MRI and PET, and CSF or blood-based biomarkers \cite{ottoy2025recent,salvado2024disease,grande2025blood}. However, some of these procedures require specialized medical resources or invasive testing, motivating the exploration of more accessible and complementary screening approaches. EEG has recently emerged as a promising alternative for the early and timely detection of AD. EEG offers several advantages, including non-invasive, real-time monitoring, affordability, and portability, making it particularly suitable for widespread clinical use~\cite{tzimourta2021machine, ieracitano2019convolutional}. Effective EEG-based AD detection may also enable integration with mobile health platforms, supporting home-based detection and continuous tracking of disease progression.

Over the past two decades, research on EEG-based AD detection has primarily followed two main directions for interpreting EEG signals. The first approach relies on handcrafted biomarker extraction, using features such as statistical~\cite{tzimourta2019eeg, tzimourta2019analysis}, spectral~\cite{wang2017enhanced, cassani2014effects}, power~\cite{fahimi2017index, schmidt2013index}, and complexity-based~\cite{garn2015quantitative, azami2019multiscale} metrics. These features are then used as input to conventional machine learning classifiers, such as multi-layer perceptrons (MLPs) and support vector machines (SVMs), for final prediction. The second direction leverages modern deep learning techniques applied to 1D EEG signals or their transformed representations. Commonly used architectures include convolutional neural networks (CNNs)~\cite{li2022predictive, ieracitano2019convolutional, cura2022deep}, graph convolutional networks (GNNs)~\cite{shan2022spatial, klepl2023adaptive}, and transformers~\cite{miltiadous2023dice}, which have demonstrated superior performance by learning spatial-temporal features.

However, existing EEG-based AD detection methods have two major limitations. First, most approaches still rely on manual feature engineering or data transformation and lack a truly end-to-end learning framework, whether they use traditional machine learning or deep learning. For instance, the methods in~\cite{bravo2024spectrocvt} and~\cite{huggins2021deep} convert 1D EEG signals into time-frequency representations using the Short-Time Fourier Transform (STFT) and Continuous Wavelet Transform (CWT), respectively, and then apply 2D CNNs to extract features from the resulting spectrogram images. While these transformations may facilitate feature learning by highlighting certain signal properties, they often distort or discard raw signal information, potentially eliminating valuable patterns in the temporal, frequency, or spatial domains. Second, since EEG-based AD datasets are limited and expensive to collect, many existing studies use relatively small cohorts, typically 20 to fewer than 200 subjects~\cite{aviles2024machine}, and evaluate models on a single dataset. This limited subject scale and demographic diversity hinder comprehensive evaluation of model robustness and generalizability, limiting these methods' clinical applicability in real-world settings. These limitations motivate the development of EEG models that learn disease-related representations directly from raw signals, capture temporal patterns at multiple scales, model spatial dependencies across channels, and evaluate performance on more diverse subject cohorts using a robust evaluation protocol.

\begin{figure}[!t]
    \centering
    \includegraphics[width=1.0\linewidth]{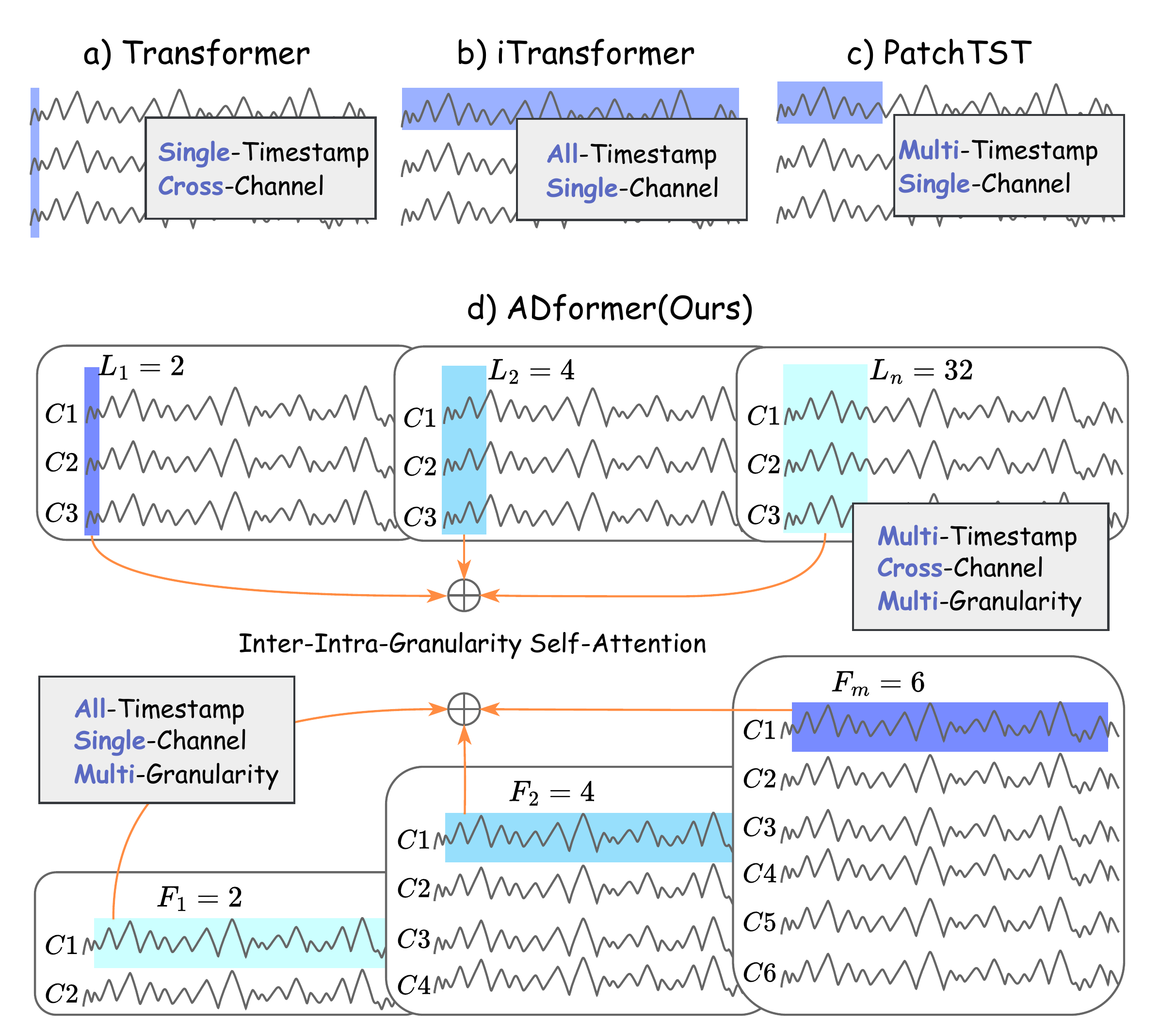}
    \caption{\textbf{Token embedding methods.} Transformer and Informer~\cite{vaswani2017attention,zhou2021informer} use a single cross-channel timestamp as a token, while iTransformer~\cite{liu2024itransformer} uses the full series from a single channel as a token. PatchTST~\cite{nie2022time} adopts a patch of timestamps from a single channel as a token. For \name, we propose a token embedding method with a branch that repeatedly segments the EEG into patches of varying lengths, each of which is defined as a granularity, and embeds these cross-channel patches. Another branch repeatedly upscales the number of EEG channels using 1-D convolutions, with each mapped pseudo-channel count defined at a given granularity, and embeds the full EEG series for each channel.
    }
    \label{fig:adformer_embedding}
\end{figure}

To bridge these gaps, we propose an end-to-end multi-granularity spatial-temporal transformer, named \textbf{\name}. Our method takes 1D EEG signals as input, eliminating the need for manual feature extraction or data transformation. We introduced a preliminary version of this work, \textbf{Medformer}, in~\cite{wang2024medformer}. It employs cross-channel, multi-granularity patch embedding and intra-inter-granularity self-attention to jointly capture spatial and temporal dependencies, and we evaluated it on various medical time-series classification tasks, including EEG and ECG. In this paper, we extend and adapt the Medformer architecture to better suit EEG signals and focus specifically on EEG-based AD detection. Given the richer spatial information in EEG signals than in other medical time series, we introduce an additional branch to model spatial features more explicitly. Specifically, we upscale the raw EEG channels to different numbers of pseudo-channels and embed each channel as a whole-series, multi-granularity channel embedding. We designate the original cross-channel multi-granularity patch embedding as the \textbf{temporal branch}, and the newly proposed whole-series multi-granularity channel embedding as the \textbf{spatial branch}. Figure~\ref{fig:adformer_embedding} provides a comparative overview of our data embedding strategy versus existing approaches used in time series transformers~\cite{vaswani2017attention, liu2024itransformer, nie2022time}. The embeddings from both branches are processed independently through intra-inter-granularity self-attention modules, enabling the model to capture local dependencies within each granularity and global interactions across granularities. We evaluate \name on four large-scale datasets comprising a total of 1,713 subjects and 2,589,094 1-second EEG samples, forming the \textbf{largest} EEG-based AD detection corpora to date. Our approach is compared against 9 baselines, including 1 traditional manual feature extraction method and 8 state-of-the-art (SOTA) deep learning models. We conduct extensive experiments under three classification settings: (1) multi-class classification across various brain disorders, (2) binary classification tasks between AD and HC subjects, and (3) three-class classification among HC, AD, and vascular dementia for differentiating various dementias. All using a rigorously designed \textbf{subject-independent} cross-validation protocol. We also conduct case studies, including the impact of sample length and overlap rate on detection performance. Our method achieves the best performance in 25 out of 36 metrics across the four datasets and three classification settings. Notably, \name achieves the best subject-level F1 score across all datasets and tasks, demonstrating strong real-world applicability. Specifically, the subject-level F1 scores for distinguishing AD from healthy controls are 87.11\%, 89.83\%, 67.99\%, and 84.42\%, respectively.

We summarize our key contributions here:
    \begin{enumerate}
        \item We introduce \name, a novel end-to-end multi-granularity spatial-temporal transformer designed explicitly for EEG-based Alzheimer's disease detection.

        \item We conduct comprehensive experiments on 4 large-scale EEG datasets, totaling 1,713 subjects and 2,589,094 1-second samples, forming the world's largest corpora to date for EEG-based AD detection and demonstrating our model's generalization capability.
        
        \item We compare \name with 1 feature-based method and 8 SOTA deep learning methods, evaluating multi-class, AD vs HC binary classification, and 3-class dementia classification tasks under a cross-validated, subject-independent setting. Case studies on sample length and overlap rate are also explored.

    \end{enumerate}

The rest of the paper is organized as follows: Section~\ref{sec:related} reviews related works, including existing methods for EEG-based AD analysis and transformers in EEG. Section~\ref{sec:preliminary} discusses the preliminary knowledge and their significance, and defines the research problem. In Section~\ref{sec:method}, we detail the proposed method's architecture. Section~\ref{sec:experimental_setups} describes the datasets used and implementation details. Section~\ref{sec:experimental_results} presents the results and their analysis for our method and all baselines under different experimental setups. Section~\ref{sec:discussion} discusses the limitations and future work of this domain. Finally, Section~\ref{sec:conclusion} concludes the paper.

\section{Related Work}
\label{sec:related}

\subsection{EEG-based Alzheimer's Disease Detection}
\label{sub:eeg_ad_detection}

In EEG-based AD analysis, two primary research directions emerge: manual feature extraction and deep learning.

\textbf{Manual Feature Extraction.}
Traditional approaches typically rely on handcrafted features that characterize EEG signals from different perspectives, followed by conventional classifiers such as Multi-Layer Perceptrons (MLPs) or Support Vector Machines (SVMs). Commonly used statistical features include mean, skewness, kurtosis, standard deviation, max, min, and median~\cite{tzimourta2019eeg, tzimourta2019analysis, kulkarni2017extracting, kanda2014clinician, waser2013eeg, tylova2013predictive, mora2019scale}. Spectral features encompass phase shift, phase coherence, bispectrum, bicoherence, spectral centroid, spectral roll-off, spectral peak, average magnitude, median frequency, and amplitude modulation~\cite{wang2017enhanced, cassani2014effects, wang2015multiple, fraga2013characterizing, tait2019network, waser2016quantifying, trambaiolli2011improving}. Power features include power spectral density, relative band power, EEG rhythm ratios, and energy~\cite{fahimi2017index, schmidt2013index, liu2016multiple, kanda2014clinician}, while complexity features comprise various entropy measures such as Shannon entropy, Tsallis entropy, dispersion entropy, and spectral entropy~\cite{garn2015quantitative, azami2019multiscale, tylova2018unbiased}. Beyond these conventional features, several studies have explored more sophisticated signal analysis techniques to characterize EEG abnormalities associated with AD. For example, the method in~\cite{amezquita2019novel} combined MUSIC-EWT decomposition with fractal dimensions and the Hurst exponent to distinguish MCI from AD. In addition, the I-FAST system introduced in~\cite{buscema2015improved} uses robust invariant features from unprocessed EEG recordings for AD diagnosis. Overall, handcrafted features offer interpretability, making them potentially useful for clinical applications.

\textbf{Deep Learning.}
In contrast to manual feature extraction, deep learning approaches automatically learn discriminative representations from EEG data, reducing the need for handcrafted features. A common strategy is to transform EEG signals into image-like representations that capture frequency- or energy-related characteristics. Examples include computing power spectral density (PSD) across channels~\cite{ieracitano2019convolutional}, generating 2D representations of EEG sub-bands (e.g., theta, alpha, beta)~\cite{ismail2019early}, and transforming EEG signals into time-frequency (TF) images using the Short-Time Fourier Transform (STFT) for 2D CNN classification~\cite{cura2022deep}.
Along this direction, the study in~\cite{morabito2016deep} explored deep CNNs for classifying MCI and AD from scalp EEG recordings. Moving beyond handcrafted feature representations, the CNN proposed in~\cite{ieracitano2020selflearning} learns directly from EEG epochs. More recently, the authors of~\cite{morabito2023explainable} investigated longitudinal MCI-to-AD analysis using PSD channel-frequency maps with a CNN, incorporating Grad-CAM to explain model predictions. Extending the focus on interpretable deep learning, EEGDecoder-X~\cite{suffian2026eegdecoder} combines 1D CNNs, channel attention, and a Transformer for cross-subject classification of AD, Creutzfeldt-Jakob disease, and healthy controls.
Beyond CNN-based methods, recent studies have also explored graph-based and attention-based architectures to capture the spatial-temporal characteristics of EEG signals. For instance, the model in~\cite{shan2022spatial} employed graph convolutional networks (GCNs) to model the topological structure of EEG connectivity and dynamic temporal information, while the model in~\cite{klepl2023adaptive} adopted gated GCNs to enhance node features with PSD similarity for adaptive graph learning. Furthermore, DICE-Net~\cite{miltiadous2023dice} integrates PSD and wavelet transforms with a 2D CNN and a self-attention module. Despite these advances, many existing deep learning methods still rely on transforming EEG signals into 2D representations, such as images or graphs, while fully end-to-end approaches that operate directly on raw 1D EEG data remain comparatively less explored.

\subsection{Transformer for EEG}
\label{sub:eeg_transformer}

End-to-end transformers have demonstrated promising results in the EEG domain across a variety of tasks. For example, the approach in~\cite{xie2022transformer} leverages spatial-temporal information to embed features using a transformer for motor imagery classification. EEGConformer, proposed in~\cite{song2022eeg}, incorporates both temporal and spatial convolutions for token embedding and has shown effectiveness on motor imagery and emotion recognition tasks. Other models, such as BIOT~\cite{yang2024biot} and Medformer~\cite{wang2024medformer}, extend transformer architectures to handle diverse biomedical signals, including EEG. These models utilize either single-channel or cross-channel patching strategies for token embedding, demonstrating strong performance across modalities. More recently, end-to-end transformers have been explored as backbone architectures for pretraining EEG foundation models. For instance, LabraM~\cite{jiang2024large}, EEGPT~\cite{wang2024eegpt}, and EEGformer~\cite{chen2024eegformer} apply techniques such as vector quantization, momentum alignment, and masked-reconstruction self-supervised learning on single-channel transformer patches. CBraMod~\cite{wang2025cbramod} integrates time-frequency patch embeddings with a two-stage transformer that separately captures temporal and spatial representations. ST-EEGFormer~\cite{yang2026eeg} explores the true effectiveness of EEG foundation models using a simple Transformer structure. These studies highlight the learning ability of end-to-end transformers for EEG classification, particularly in scalability and representation learning. This motivates further exploration of end-to-end transformer architectures for EEG-based AD detection, a domain where such methods remain underexplored.

\section{Preliminaries and Problem Definition}
\label{sec:preliminary}

\subsection{Dementia and Alzheimer's Disease}
\label{sub:dementia_ad}

Dementia is a clinical syndrome characterized by cognitive decline that interferes with daily functioning, rather than a single disease. Alzheimer's disease is its most common underlying cause, accounting for approximately 60-80\% of dementia cases \cite{alzheimer20152015,knopman2021alzheimer}. Other major causes include vascular dementia (VD), frontotemporal dementia (FTD), and dementia with Lewy bodies (LBD), which may share similar clinical symptoms but differ in their underlying pathological mechanisms \cite{scheltens2016alzheimer,jack2024revised}. For instance, AD is characterized by amyloid-$\beta$ plaque accumulation and tau neurofibrillary tangles, whereas VD is primarily associated with cerebrovascular damage, such as stroke or cerebral small-vessel disease. In this study, we follow the diagnostic labels in the original EEG datasets and focus primarily on AD detection.

\subsection{EEG-Based Disease Detection}
\label{sub:eeg_disease_detection}
In EEG data collected for disease detection tasks, each subject is typically assigned a single label indicating their brain health status, such as the presence or absence of specific neurological conditions. In some cases, subjects may receive multiple labels if they have comorbid neurological conditions, since many brain diseases are not mutually exclusive~\cite{kim2023deep}.
To improve memory and computational efficiency in deep learning models, long EEG recordings (e.g., trials or sessions) are commonly segmented into shorter samples. As a result, each EEG sample is typically associated with a disease label and a subject ID. To determine whether a subject has a particular disease, majority-voting post-processing techniques often aggregate predictions across all samples from the same subject, assigning the subject the label that appears most frequently among their samples~\cite{ieracitano2019convolutional}.

\subsection{Subject-Independent Evaluation}
\label{sub:sub_indep}
\begin{figure}[!t]  
    \centering
    \includegraphics[width=1.0\linewidth]{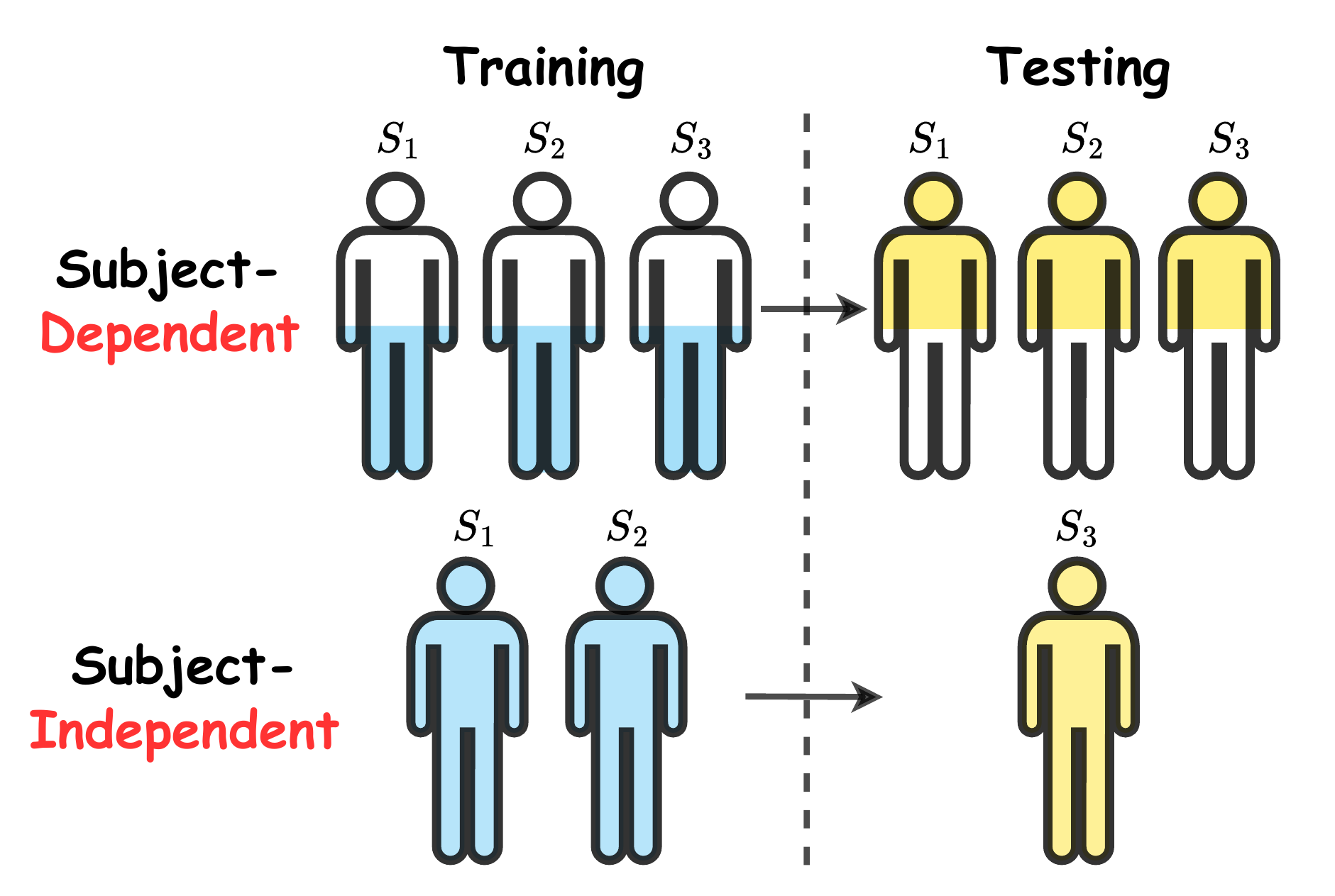}
    \caption{\textbf{Subject-dependent/independent setups.} 
    In the subject-dependent setup, samples from the same subject can appear in both the training and test sets, causing information leakage and allowing shortcut learning~\cite{wang2024causes}. In the subject-independent setup, samples from the same subject are exclusively in either the training or test set, which is more challenging and practically meaningful.
    }
    \label{fig:subject-dependent-independent}
\end{figure}

In EEG-based brain disease detection, subject-specific features can significantly influence model performance. As a result, different experimental setups may yield vastly different outcomes, potentially leading to misleading conclusions~\cite{kunjan2021necessity}. Therefore, experimental protocols must be carefully designed to align with real-world clinical applications. In the context of EEG-based AD detection, two commonly used evaluation setups exist: subject-dependent\cite{nour2024novel, kumar2023eegalzheimer} and subject-independent\cite{watanabe2024deep, chen2024multi}. 
In subject-dependent setups, samples from all subjects are randomly shuffled and then split into training, validation, and test sets. As a result, data from the same subject may appear in multiple sets, potentially causing information leakage and overestimating model performance. In contrast, subject-independent setups split the data strictly by subject, ensuring that samples from the same subject appear in only one of the three subsets: training, validation, or testing. Figure~\ref{fig:subject-dependent-independent}, adapted from~\cite{wang2024contrast}, illustrates these two setups. In this work, we adopt the subject-independent setup, which better simulates real-world deployment scenarios with unseen subjects and prevents information leakage from subject-specific features during training.

\subsection{Problem Definition}
\label{sub:problem_definition}
\textbf{EEG-based Alzheimer's Disease Detection.}
Let the input EEG sample be denoted as $\bm{x}_{\text{in}} \in \mathbb{R}^{T \times C}$, where $T$ is the number of time points and $C$ is the number of EEG channels. The objective is to learn a meaningful representation $\bm{h}$ to predict the corresponding label $\bm{y} \in \mathbb{R}^{K}$. Here, $K$ denotes the number of target classes for neurological conditions, including HC, MCI, AD, FTD, and other brain disorders. In this study, we primarily focus on AD detection.

\section{Method}
\label{sec:method}

\begin{figure*}[!t]
    \centering
    \includegraphics[width=1.0\linewidth]{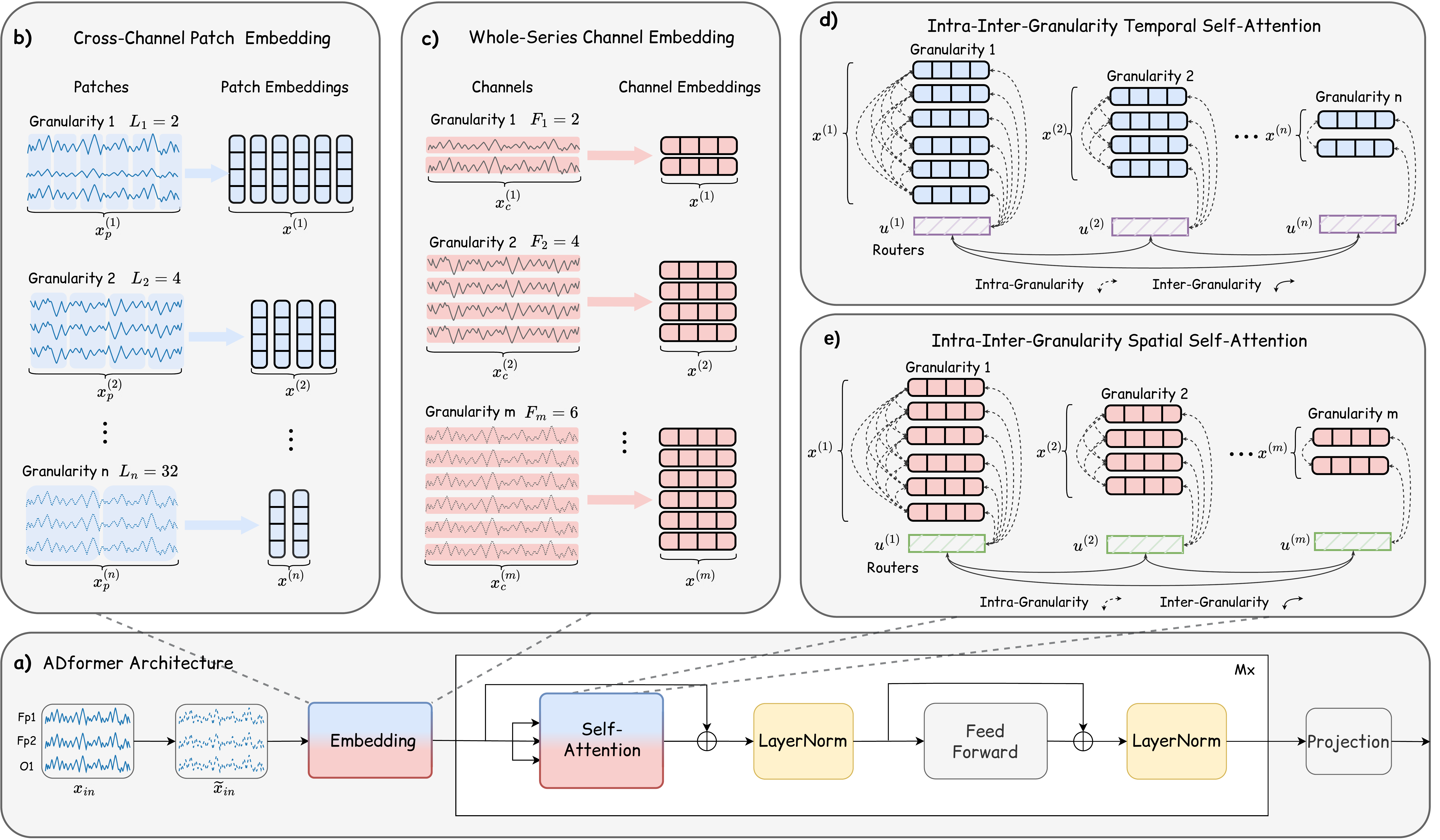}
    \caption{\textbf{Overview of \name.} 
    \textbf{a)} Architecture.
    \textbf{b)} For the $i$-th patch length $L_i$ (granularity $i$) and input sample $\bm{x}_{\text{in}}$, we first apply data augmentation to obtain $\widetilde{\bm{x}}_{\text{in}}^{(i)}$, which is then segmented into cross-channel non-overlapping patches $\bm{x}_{\text{p}}^{(i)}$. These patches are projected into the latent space and fused with granularity and positional embeddings to yield the final patch representation $\bm{x}^{(i)}$.
    \textbf{c)} For the $j$-th pseudo-channel number $F_j$ (granularity $j$) and input $\bm{x}_{\text{in}}$, we generate an augmented view $\widetilde{\bm{x}}_{\text{in}}^{(j)}$. Channel-wise positional embeddings are added to the transposed input $(\widetilde{\bm{x}}_{\text{in}}^{(j)})^{\mathsf{T}}$, followed by a linear projection to scale the channel dimension. We then project the resulting channel series into the latent space and combine it with granularity embeddings to produce the final channel representation $\bm{x}^{(j)}$.
    \textbf{d-e)} We perform intra-granularity self-attention to model dependencies within each granularity and inter-granularity self-attention to enable cross-granularity interaction. We introduce a router mechanism in the inter-granularity attention to reduce computational and memory complexity.
    }
    \label{fig:adformer_framework}
\end{figure*}

\subsection{Overview}
\label{sub:method_overview}

Our method incorporates temporal and spatial branches to capture features across two dimensions. The temporal branch uses cross-channel multi-granularity patch embedding, which embeds patches of varying lengths at different granularities as self-attention tokens. The spatial branch applies whole-series multi-granularity channel embedding, first scaling up the number of channels across granularities, then embedding each pseudo-channel as tokens for self-attention. In both branches, we perform intra-inter-granularity self-attention in two stages. First, intra-granularity self-attention captures local features within each granularity. Then, inter-granularity self-attention, implemented via a router mechanism, learns interactions across granularities. The two branches operate in parallel. Figure~\ref{fig:adformer_framework} illustrates the overall architecture.

\subsection{Cross-Channel Multi-Granularity Patch Embedding}
\label{sub:patch_embedding}

Drawing on neuroscience and brain connectivity, we assume that inherent correlations exist among different channels in EEG data~\cite{bazinet2023towards}. Motivated by this assumption, we propose cross-channel patching for token embedding that captures both temporal and spatial features in a single step. Additionally, EEG biomarkers embedded within specific frequency bands~\cite{fahimi2017index,liu2016multiple} further motivate adopting multi-granularity embedding. Unlike traditional techniques such as up/downsampling or handcrafted band filtering, multi-granularity learning enables the model to capture EEG patterns at different temporal scales.

Given these motivations, we propose a novel token embedding strategy: cross-channel multi-granularity patch embedding. Area b) in Figure~\ref{fig:adformer_framework} shows the steps for embeddings. Given a multivariate EEG input sample $\bm{x}_{\text{in}} \in \mathbb{R}^{T \times C}$ and a set of patch lengths $\left\{ {L}_{1},{L}_{2},\ldots,{L}_{n} \right\}$, for the $i$-th patch length $L_i$ (corresponding to granularity $i$), we first apply data augmentations to generate augmented views $\widetilde{\bm{x}}_{\text{in}}^{(i)} \in \mathbb{R}^{T \times C}$. Inspired by the contrastive learning framework~\cite{chen2020simple}, we argue that these augmentations enhance the subsequent inter-granularity self-attention stage by promoting knowledge sharing across granularities. We adopt the same augmentation strategy as Medformer~\cite{wang2024medformer}. Next, we segment the augmented view $\widetilde{\bm{x}}_{\text{in}}^{(i)}$ into $N_{i}$ non-overlapping cross-channel patches $\bm{x}_{\text{p}}^{(i)} \in \mathbb{R}^{N_{i} \times \left( L_{i} \cdot C \right)}$. We apply zero padding to ensure the total number of time steps $T$ is divisible by $L_i$, yielding $N_{i} = \left\lceil {T/L_{i}} \right\rceil$. Each patch is then projected into a latent embedding space via a linear transformation: 

\begin{equation}
    \bm{x}_{\text{e}}^{(i)} = \bm{x}_{\text{p}}^{(i)}\bm{W}^{(i)}, 
\end{equation}

where $\bm{x}_{\text{e}}^{(i)} \in \mathbb{R}^{N_{i} \times D}$ and $\bm{W}^{(i)} \in \mathbb{R}^{\left( L_{i} \cdot C \right) \times D}$.

A fixed positional embedding matrix $\bm{W}_{\text{pos}} \in \mathbb{R}^{G \times D}$ is employed for positional encoding~\cite{vaswani2017attention}, where $G$ is a sufficiently large constant. We add the first $N_i$ rows of this matrix, $\bm{W}_{\text{pos}}[1:N_i] \in \mathbb{R}^{N_{i} \times D}$, along with a learnable granularity embedding $\bm{W}_{\text{gr}}^{(i)} \in \mathbb{R}^{1 \times D}$ for the $i$-th patch length $L_i$, to form the final patch embedding for granularity $i$:

\begin{equation}
    \bm{x}^{(i)} = \bm{x}_{\text{e}}^{(i)} + \bm{W}_{\text{pos}}[1:N_i] + \bm{W}_{\text{gr}}^{(i)},
\end{equation}

where $\bm{x}^{(i)} \in {\mathbb{R}}^{N_{i} \times D}$. Note that $\bm{W}_{\text{gr}}^{(i)}$ is broadcasted to all $N_i$ patch embeddings during addition.

To facilitate interaction across granularities and to reduce computational complexity in subsequent intra-inter-granularity self-attention layers, we also introduce a router token here:

\begin{equation}
    \bm{u}^{(i)} = \bm{W}_{\text{pos}}[N_i+1] + \bm{W}_{\text{gr}}^{(i)},
\end{equation}

where $\bm{u}^{(i)}, \bm{W}_{\text{pos}}[N_i+1], \bm{W}_{\text{gr}}^{(i)} \in \mathbb{R}^{1 \times D}$. Here, $\bm{W}_{\text{pos}}[N_i+1]$ is not used for positional encoding but serves to inform the number of patches at granularity $L_i$, while $\bm{W}_{\text{gr}}^{(i)}$ conveys granularity-specific information. Together, these components distinguish routers associated with different granularities.

In the end, we obtain a set of patch $\left\{ \bm{x}^{(1)},\bm{x}^{(2)},\ldots,\bm{x}^{(n)} \right\}$ and router $\left\{ \bm{u}^{(1)},\bm{u}^{(2)},\ldots,\bm{u}^{(n)} \right\}$ embeddings corresponding to the patch lengths $\left\{ L_{1}, L_{2}, \ldots, L_{n} \right\}$, collectively denoting multi-granularities. They are fed as input tokens into the subsequent two-stage intra-inter-granularity self-attention module.

\subsection{Whole-Series Multi-Granularity Channel Embedding}
\label{sub:channel_embedding}

Compared with other medical time series, such as ECG and EMG, EEG signals typically have more channels due to the electrode system, enabling finer spatial feature representation. To leverage this, we add a spatial branch, extending \textit{Medformer}~\cite{wang2024medformer} to better capture EEG-specific spatial features. Unlike the temporal branch, this spatial branch embeds the entire time series for each channel as a single token. For multi-granularity learning, we upscale the number of channels to generate pseudo-channels, where more pseudo-channels correspond to finer feature utilization and thus finer granularity.

We propose a novel token embedding method, whole-series multi-granularity channel embedding, to improve data representation and capture complex channel interactions across granularity levels. Area c) in Figure~\ref{fig:adformer_framework} shows the steps for embeddings. Given an input multivariate EEG sample $\bm{x}_{\text{in}} \in \mathbb{R}^{T \times C}$, and a list of scaled channel number $\left\{ {F}_{1},{F}_{2},\ldots,{F}_{m} \right\}$. For the $j$-th scaled channel number $F_j$ denoting granularity $j$, we first apply data augmentations to obtain augmented input $\widetilde{\bm{x}}_{\text{in}}^{(j)} \in \mathbb{R}^{T \times C}$. Again, the data augmentation is intended to enhance learning during the later inter-granularity self-attention.

Next, we transpose the input sample and add a learnable channel-wise positional embedding $\bm{W}_{\text{pos}}[1:C]$ to obtain: 

\begin{equation}
    \bm{x}_{\text{trans}} = (\widetilde{\bm{x}}_{\text{in}}^{(j)})^{\mathsf{T}} + \bm{W}_{\text{pos}}[1:C], 
\end{equation}

where $\bm{x}_{\text{trans}}, (\widetilde{\bm{x}}_{\text{in}}^{(j)})^{\mathsf{T}}, \bm{W}_{\text{pos}}[1:C] \in \mathbb{R}^{C \times T}$. Unlike the temporal branch, we add the positional embedding $\bm{W}_{\text{pos}}[1:C]$ directly to the transposed raw input, since the subsequent channel scaling-up process breaks the original channel order. 

We then scale up the channel number with a linear projection $\bm{W}_{1}^{(j)}$ to generate pseudo-channels for finer spatial feature extraction. For pseudo-channel number $F_j$ denoting granularity $j$, we obtain $\bm{x}_{\text{c}}^{j} =\bm{W}_{1}^{(j)} \bm{x}_{\text{trans}}$, where $\bm{x}_{\text{c}}^{j} \in \mathbb{R}^{F_{j} \times T}$ and $\bm{W}_{1}^{(j)} \in \mathbb{R}^{F_{j} \times C}$. The result $\bm{x}_{\text{c}}^{j}$ is further projected into a latent embedding space using another linear projection: $\bm{x}_{\text{e}}^{(j)} = \bm{x}_{\text{c}}^{(j)}\bm{W}_{2}^{(j)}$, where $\bm{x}_{\text{e}}^{(j)} \in \mathbb{R}^{F_{j} \times D}$ and $\bm{W}_{2}^{(j)} \in \mathbb{R}^{T \times D}$. Since positional embedding has already been added, we only add a learnable granularity embedding $\bm{W}_{\text{gr}}^{(j)} \in \mathbb{R}^{1 \times D}$ to help distinguish the $j$-th granularity, resulting in the final channel embedding:

\begin{equation}
    \bm{x}^{(j)} = \bm{x}_{\text{e}}^{(j)} + \bm{W}_{\text{gr}}^{(j)},
\end{equation}

where $\bm{x}^{(j)} \in {\mathbb{R}}^{F_j \times D}$. We also enable a router token here, but only use the granularity embedding $\bm{W}_{\text{gr}}^{(j)}$: 

\begin{equation}
    \bm{u}^{(j)} = \bm{W}_{\text{gr}}^{(j)},
\end{equation}

Finally, we obtain a list of channel $\left\{ \bm{x}^{(1)},\bm{x}^{(2)},\ldots,\bm{x}^{(m)} \right\}$ and router  $\left\{ \bm{u}^{(1)},\bm{u}^{(2)},\ldots,\bm{u}^{(m)} \right\}$ embeddings of different scaled channel numbers $\left\{ F_{1}, F_{2}, \ldots, F_{m} \right\}$, collectively denoting multi-granularities. They are fed as input tokens into the later two-stage self-attention module.

\subsection{Intra-Inter-Granularity Self-Attention}
\label{sub:intra_inter_attention}

Our goal is to learn multi-granularity features for both the temporal and spatial branch embeddings, while enabling interaction across granularities during self-attention. Taking the temporal branch as an example, a naive approach would be to concatenate all patch embeddings $\left\{ \bm{x}^{(1)}, \bm{x}^{(2)}, \ldots, \bm{x}^{(n)} \right\}$ into a large patch embedding $\bm{X} \in {\mathbb{R}}^{\left( \sum_{i=1}^n N_i \right) \times D}$ and perform self-attention on this new embedding, where $n$ denotes the number of different granularities. However, this approach leads to a time complexity of $O\left(\left( \sum_{i=1}^n N_i \right)^2 \right)$, which becomes computationally infeasible for large $n$.

To reduce the computational complexity, we introduce a router mechanism and divide the self-attention process into two distinct stages: a) intra-granularity and b) inter-granularity self-attention. The intra-granularity stage performs self-attention within each individual granularity to capture its unique features. In contrast, the inter-granularity stage performs self-attention across different granularities, enabling the model to learn shared and complementary features across scales. \textit{The intra-inter-granularity self-attention mechanisms for the temporal and spatial branches operate independently and do not share parameters, though they follow the same procedural steps}. Parts d) and e) in Figure~\ref{fig:adformer_framework} illustrate the intra- and inter-granularity processes for the temporal and spatial branches, respectively.

\subsubsection{Intra-Granularity Self-Attention}
\label{subsub:intra_attention}

For granularity $i$, we concatenate the token embedding $\bm{x}^{(i)}\in {\mathbb{R}}^{N_{i} \times D}$ and router embedding $\bm{u}^{(i)}\in {\mathbb{R}}^{1 \times D}$ to form an intermediate sequence of
embeddings $\bm{z}^{(i)} \in {\mathbb{R}}^{\left( N_{i} + 1 \right) \times D}$:

\begin{equation}
    \begin{aligned}
        \bm{z}^{(i)} = & \left\lbrack \bm{x}^{(i)}\|\bm{u}^{(i)} \right\rbrack \\
    \end{aligned}
\end{equation}

where $\left\lbrack \cdot \| \cdot \right\rbrack$ denotes concatenation. Self-attention is then performed on $\bm{z}^{(i)}$ for both the token and router embeddings:

\begin{equation}
    \begin{aligned}
        \bm{x}^{(i)} \leftarrow & \operatorname{ Attn^{Intra}}\left( \bm{x}^{(i)},\bm{z}^{(i)},\bm{z}^{(i)} \right) \\
        \bm{u}^{(i)} \leftarrow & \operatorname{ Attn^{Intra}}\left( \bm{u}^{(i)},\bm{z}^{(i)},\bm{z}^{(i)} \right) \\
    \end{aligned}
\end{equation}

where $\operatorname{ Attn^{Intra}}\left( \bm{Q}, \bm{K}, \bm{V} \right)$ refers to the scaled dot-product attention mechanism described in~\cite{vaswani2017attention}. The router embedding $\bm{u}^{(i)}$ is updated simultaneously with the token embedding $\bm{x}^{(i)}$ to ensure consistency. This update lets the router embedding summarize features at the current training step at the same granularity, while the token embeddings receive global information from the router. The intra-granularity self-attention mechanism helps the model to capture features within the same granularity, facilitating the extraction of local features and correlations among timestamps of the same scale.

\begin{table*}[!t]
    \centering
    \caption{\textbf{Processed Dataset Statistics.} All datasets are downsampled to 128Hz, segmented into 1s, half overlapping samples. \textbf{RS} - Resting State; \textbf{PS} - Photic Stimulation; \textbf{HV} - Hyperventilation. \textbf{HC} - Healthy Controls; \textbf{AD}: Alzheimer’s Disease; \textbf{FTD}: Frontotemporal Dementia; \textbf{MCI}: Mild Cognitive Impairment; \textbf{SCZ}: Schizophrenia; \textbf{DEP}: Depression; \textbf{DEM}: Dementia. 
    }
    \label{tab:processed_data}
    \resizebox{1.0\textwidth}{!}{%
    \begin{tabular}{@{}l|cccccc@{}}
    \toprule
    \multicolumn{1}{l|}{\textbf{Datasets}}  & \textbf{Task} & \textbf{\#Subjects} & \textbf{Sampling Rate} & \textbf{\#Channels} & \textbf{\#1-s Samples} & \textbf{Classes}  \\ 
    \midrule
    \multicolumn{1}{l|}{\textbf{ADFTD}} & RS \& PS  & 88 &  500-->128 Hz  & 19 &  185,656  & HC vs AD vs FTD \\
    \multicolumn{1}{l|}{\textbf{CNBPM}} & RS  & 189 &  256-->128 Hz  & 19 & 140,165  & HC vs AD vs MCI \\
    \multicolumn{1}{l|}{\textbf{P-ADIC}} & RS  & 249 &  500-->128 Hz  & 19 &  542,490  & HC vs AD vs MCI vs SCZ vs DEP \\
    \multicolumn{1}{l|}{\textbf{CAUEEG}} & RS \& PS \& HV & 1187 &  200-->128 Hz  & 19 &  1,720,783 & HC vs DEM vs MCI \\
    
    \bottomrule
    \end{tabular}
    } 
\end{table*}

\subsubsection{Inter-Granularity Self-Attention}
\label{subsub:inter_attention}

We concatenate all router embeddings $\left\{ \bm{u}^{(1)},\bm{u}^{(2)},\ldots,\bm{u}^{(n)} \right\}$ to form a sequence of routers $\bm{U} \in {\mathbb{R}}^{n \times D}$:

\begin{equation}
        \bm{U} = \left\lbrack \bm{u}^{(1)}\|\bm{u}^{(2)}\|\ldots\|\bm{u}^{(n)} \right\rbrack \\
\end{equation}

where $n$ is the number of different granularities. For each granularity $i$ with patch length $L_i$, we apply self-attention to the router embedding $\bm{u}^{(i)} \in {\mathbb{R}}^{1 \times D}$ with all the routers $\bm{U}$:

\begin{equation}
    \begin{aligned}
        \bm{u}^{(i)} \leftarrow & \operatorname{ Attn^{Inter}}\left( \bm{u}^{(i)},\bm{U},\bm{U} \right) \\
    \end{aligned}
\end{equation}

This process enables each router, which already incorporates local features via intra-granularity self-attention, to exchange information with other routers of different granularities. Thus, inter-granularity self-attention effectively captures cross-granularity dependencies and interactions across various levels. Moreover, the router mechanism significantly reduces the time complexity compared to the naive approach. For the temporal branch, for example, the complexity drops from $O\left(\left( \sum_{i=1}^n N_i \right)^2 \right)$ to $O\left(\sum_{i=1}^n N_i^2 + n^2 \right)$. Given that $N_i \leq T$, the worst-case time complexity for the intra-inter-granularity temporal self-attention becomes $O\left( nT^2 + n^2\right)$, offering a more efficient solution for multi-granularity modeling.

\subsection{Classifier}
\label{sub:classifier}
For a given EEG input sample $\bm{x}_{\textrm{in}}$, after $M$ layers of encoding, we obtain a list of updated routers for the temporal branch ${\bm{u}^{(1)}, \bm{u}^{(2)}, \ldots, \bm{u}^{(n)}}$ and the spatial branch ${\bm{u}^{(1)}, \bm{u}^{(2)}, \ldots, \bm{u}^{(m)}}$, where $n$ and $m$ denote the number of granularities for the temporal and spatial branches, respectively. These router embeddings are concatenated and flattened to form a final representation $\bm{h} \in {\mathbb{R}}^{(n+m) \times D}$, which is then passed through a linear classifier to predict the label $\bm{\hat{y}} \in \mathbb{R}^{K}$. The model is trained using the cross-entropy loss computed with the ground-truth $\bm{y} \in \mathbb{R}^{K}$.

\section{Experimental Setups}
\label{sec:experimental_setups}

\subsection{Datasets}
\label{sub:datasets}

To demonstrate our method's end-to-end learning capability for EEG-based AD detection, we use four widely used, large-scale datasets that include subjects diagnosed with AD, other brain disorders, and healthy controls. These datasets include EEG recording tasks such as resting-state, photic stimulation, and hyperventilation. The datasets used are: \textbf{ADFTD}~\cite{miltiadous2023dataset}, \textbf{CNBPM}~\cite{amezquita2019novel}, \textbf{P-ADIC}~\cite{shor2021eeg}\footnote{\url{https://datadryad.org/dataset/doi:10.5061/dryad.8gtht76pw}}, and \textbf{CAUEEG}~\cite{kim2023deep}\footnote{\url{https://github.com/ipis-mjkim/caueeg-dataset}}, comprising a total of \textit{1,713 subjects}. The photic stimulation sub-dataset\footnote{\url{https://openneuro.org/datasets/ds006036/versions/1.1.0}} of ADFTD must be version 1.1.0 or later, as the older version has a subject ID mismatch with the resting-state sub-dataset\footnote{\url{https://openneuro.org/datasets/ds004504/versions/1.0.9}} and could cause information leakage when used together. All subjects provided informed consent, as stated in the README files for the datasets. We remove artifacts using Independent Component Analysis (ICA) combined with ICLabel~\cite{pion2019iclabel}, which automatically identifies and eliminates components associated with artifacts such as eye blinks, muscle activity, and heartbeats. We apply a band-pass filter between 0.5 Hz and 45 Hz and resample all signals to 128 Hz. We segment each EEG trial into 1-second samples with 50\% overlap. Z-score normalization is applied to each sample on each channel. This results in a total of \textit{2,589,094} processed 1-s samples. Table~\ref{tab:processed_data} presents detailed dataset statistics. More details on per-dataset pre-processing are available in our GitHub repository.

\subsection{Baselines}
\label{sub:baselines}
We compare our method against 9 baselines, including 1 traditional feature-based approach and 8 deep learning models. The feature-based method combines commonly used \textbf{Statistical}~\cite{tzimourta2019eeg, tzimourta2019analysis}, \textbf{Spectral}~\cite{wang2017enhanced, cassani2014effects}, \textbf{Power}~\cite{fahimi2017index, schmidt2013index}, and \textbf{Complexity}~\cite{garn2015quantitative, azami2019multiscale} features (in total 32) for EEG-based AD detection, followed by classification using a linear classifier. The deep learning baselines are either SOTA models or have demonstrated strong performance in EEG and medical time series classification tasks. The 8 deep learning models include: \textbf{EEGNet}~\cite{lawhern2018eegnet}, \textbf{TST}~\cite{zerveas2021transformer}, \textbf{EEGInception}~\cite{zhang2021eeg}, \textbf{EEGConformer}~\cite{song2022eeg}, \textbf{BIOT}~\cite{yang2024biot}, \textbf{Medformer}~\cite{wang2024medformer}, \textbf{MedGNN}~\cite{fan2025towards}, and \textbf{ModernTCN}~\cite{luo2024moderntcn}.

\begin{table*}[!t]
    \centering
    \caption{\textbf{Multi-Class Classification Results.} All classes in the datasets are used for brain disorder detection.
    }
    \label{tab:multi_class_tasks}
    \resizebox{\textwidth}{!}{
    \begin{tabular}{@{}ll|cc|cc|cc|cc@{}}
    \toprule

    \multicolumn{2}{l|}{\textbf{Datasets}}
    & \multicolumn{2}{c|}{\makecell{\textbf{ADFTD} \\ \textit{(185,656 Samples)}  \\ \textit{(88 Subjects, 3 Classes)} }}
    & \multicolumn{2}{c|}{\makecell{\textbf{CNBPM}  \\ \textit{(140,165 Samples)}  \\ \textit{(189 Subjects, 3 Classes)}  }}
    & \multicolumn{2}{c|}{\makecell{\textbf{P-ADIC} \\ \textit{(542,490  Samples)}  \\ \textit{(249 Subjects, 5 Classes)} \\ }}
    & \multicolumn{2}{c}{\makecell{\textbf{CAUEEG} \\ \textit{(1,720,783 Samples)}  \\ \textit{(1187 Subjects, 3 Classes)} \\ }}
    \\ \midrule

    \multicolumn{2}{l|}{\diagbox{\textbf{Methods}}{\textbf{Metrics}}} & \textbf{Accuracy} & \textbf{F1 Score} & \textbf{Accuracy} & \textbf{F1 Score} & \textbf{Accuracy} & \textbf{F1 Score} & \textbf{Accuracy} & \textbf{F1 Score} \\ \midrule

    \multicolumn{10}{c}{\textbf{Sample-Level Classification}}  \\
    \midrule

    \multicolumn{2}{l|}{\textbf{ManualFeature}}  & 43.37\std{3.15} & 40.65\std{3.07} & 53.53\std{5.67} & 47.83\std{5.53} & 33.25\std{3.34} & 24.25\std{1.82} & 43.84\std{2.03} & 42.60\std{2.25}  \\
    \multicolumn{2}{l|}{\textbf{EEGNet}}  & 46.86\std{2.85} & 39.36\std{2.78} & 55.27\std{7.94} & 47.03\std{4.80} & \textbf{38.78\std{1.78}} & 19.20\std{2.11} & 45.72\std{0.63} & 45.26\std{0.69}  \\
    \multicolumn{2}{l|}{\textbf{TST}}  & 48.79\std{2.93} & 44.89\std{3.20} & 64.38\std{5.28} & 60.45\std{5.44} & 32.25\std{3.68} & 25.90\std{3.01} & 49.08\std{1.35} & 48.90\std{1.56}  \\
    \multicolumn{2}{l|}{\textbf{EEGInception}}  & 50.80\std{2.48} & \textbf{48.15\std{1.98}} & 63.74\std{3.58} & 59.56\std{3.67} & 32.01\std{4.54} & 25.63\std{4.05} & 50.60\std{1.35} & 49.85\std{1.88}  \\
    \multicolumn{2}{l|}{\textbf{EEGConformer}}  & 51.23\std{3.51} & 47.47\std{3.76} & 65.99\std{4.79} & 61.97\std{4.90} & 33.05\std{3.85} & 26.39\std{2.94} & \textbf{51.41\std{1.31}} & \textbf{51.21\std{1.55}}  \\
    \multicolumn{2}{l|}{\textbf{BIOT}}  & 49.69\std{5.44} & 45.51\std{4.88} & 63.88\std{5.67} & 59.09\std{4.84} & 33.12\std{3.46} & 25.35\std{2.55} & 48.02\std{1.36} & 47.77\std{1.43}  \\
    \multicolumn{2}{l|}{\textbf{MedGNN}}  & 49.74\std{3.87} & 46.23\std{4.14} & 67.38\std{3.70} & 63.09\std{3.31} & 34.44\std{2.63} & 27.24\std{3.15} & 49.82\std{1.31} & 49.63\std{1.46}  \\
    \multicolumn{2}{l|}{\textbf{Medformer}}  & 52.58\std{1.60} & 46.49\std{3.48} & 66.36\std{4.22} & 62.30\std{4.21} & 35.07\std{3.93} & 27.37\std{2.35} & 49.40\std{0.90} & 48.80\std{1.14}  \\
    \multicolumn{2}{l|}{\textbf{ModernTCN}}  & 49.49\std{5.71} & 46.24\std{4.23} & 61.44\std{5.88} & 56.22\std{6.08} & 35.14\std{5.08} & 25.40\std{2.11} & 49.87\std{1.59} & 49.96\std{1.59}  \\
    \multicolumn{2}{l|}{\textbf{ADformer (Ours)}}  & \textbf{53.21\std{4.25}} & 46.91\std{4.33} & \textbf{67.92\std{3.78}} & \textbf{63.51\std{3.52}} & 35.36\std{3.31} & \textbf{28.10\std{1.63}} & 50.49\std{0.96} & 50.06\std{1.23}  \\

    \midrule

    \multicolumn{10}{c}{\textbf{Subject-Level Detection}}  \\
    \midrule

    \multicolumn{2}{l|}{\textbf{ManualFeature}}  & 52.63\std{8.81} & 45.13\std{11.08} & 55.90\std{6.96} & 51.75\std{7.74} & 39.23\std{2.61} & 21.68\std{3.72} & 50.04\std{4.80} & 45.62\std{6.27}  \\
    \multicolumn{2}{l|}{\textbf{EEGNet}}  & 52.63\std{5.77} & 43.15\std{6.75} & 56.41\std{1.62} & 52.70\std{3.97} & 41.54\std{3.12} & 18.81\std{2.84} & 50.79\std{0.86} & 48.66\std{0.85}  \\
    \multicolumn{2}{l|}{\textbf{TST}}  & 56.84\std{7.74} & 50.22\std{9.12} & 58.46\std{9.37} & 58.39\std{9.10} & 40.00\std{5.22} & 31.28\std{4.10} & 58.41\std{1.99} & 57.08\std{2.64}  \\
    \multicolumn{2}{l|}{\textbf{EEGInception}}  & 54.74\std{4.21} & 50.77\std{4.31} & 60.51\std{6.80} & 60.09\std{6.57} & 33.46\std{5.10} & 24.83\std{6.52} & 59.16\std{1.69} & 56.49\std{3.51}  \\
    \multicolumn{2}{l|}{\textbf{EEGConformer}}  & 58.95\std{6.98} & 53.02\std{9.41} & 63.59\std{2.99} & 63.58\std{3.03} & 36.92\std{6.37} & 26.76\std{3.95} & 59.75\std{2.24} & 58.54\std{2.17}  \\
    \multicolumn{2}{l|}{\textbf{BIOT}}  & 55.79\std{11.82} & 50.06\std{12.63} & 61.03\std{2.99} & 60.21\std{3.38} & 40.77\std{6.48} & 28.01\std{7.45} & 57.74\std{3.13} & 56.52\std{3.64}  \\
    \multicolumn{2}{l|}{\textbf{MedGNN}}  & 54.74\std{7.88} & 50.10\std{8.92} & 64.62\std{2.99} & 64.31\std{3.21} & 38.46\std{4.03} & 26.99\std{7.41} & 57.32\std{1.61} & 55.99\std{1.64}  \\
    \multicolumn{2}{l|}{\textbf{Medformer}}  & 58.95\std{3.94} & 51.60\std{6.62} & 62.05\std{6.96} & 62.00\std{6.94} & 40.38\std{3.44} & 29.16\std{0.80} & 58.58\std{1.09} & 56.45\std{1.50}  \\
    \multicolumn{2}{l|}{\textbf{ModernTCN}}  & 54.74\std{7.14} & 51.88\std{7.14} & 55.38\std{6.99} & 53.97\std{7.92} & 38.85\std{7.13} & 23.73\std{6.04} & 59.25\std{1.56} & 58.51\std{1.77}  \\
    \multicolumn{2}{l|}{\textbf{ADformer (Ours)}}  & \textbf{61.05\std{7.14}} & \textbf{53.78\std{7.99}} & \textbf{65.64\std{4.76}} & \textbf{65.88\std{4.77}} & \textbf{42.69\std{4.77}} & \textbf{32.07\std{3.48}} & \textbf{60.75\std{1.14}} & \textbf{59.20\std{1.57}}  \\

    \bottomrule
    \end{tabular}
    }
\end{table*}

\begin{figure*}[!t]
    \centering
    \includegraphics[width=1.0\linewidth]{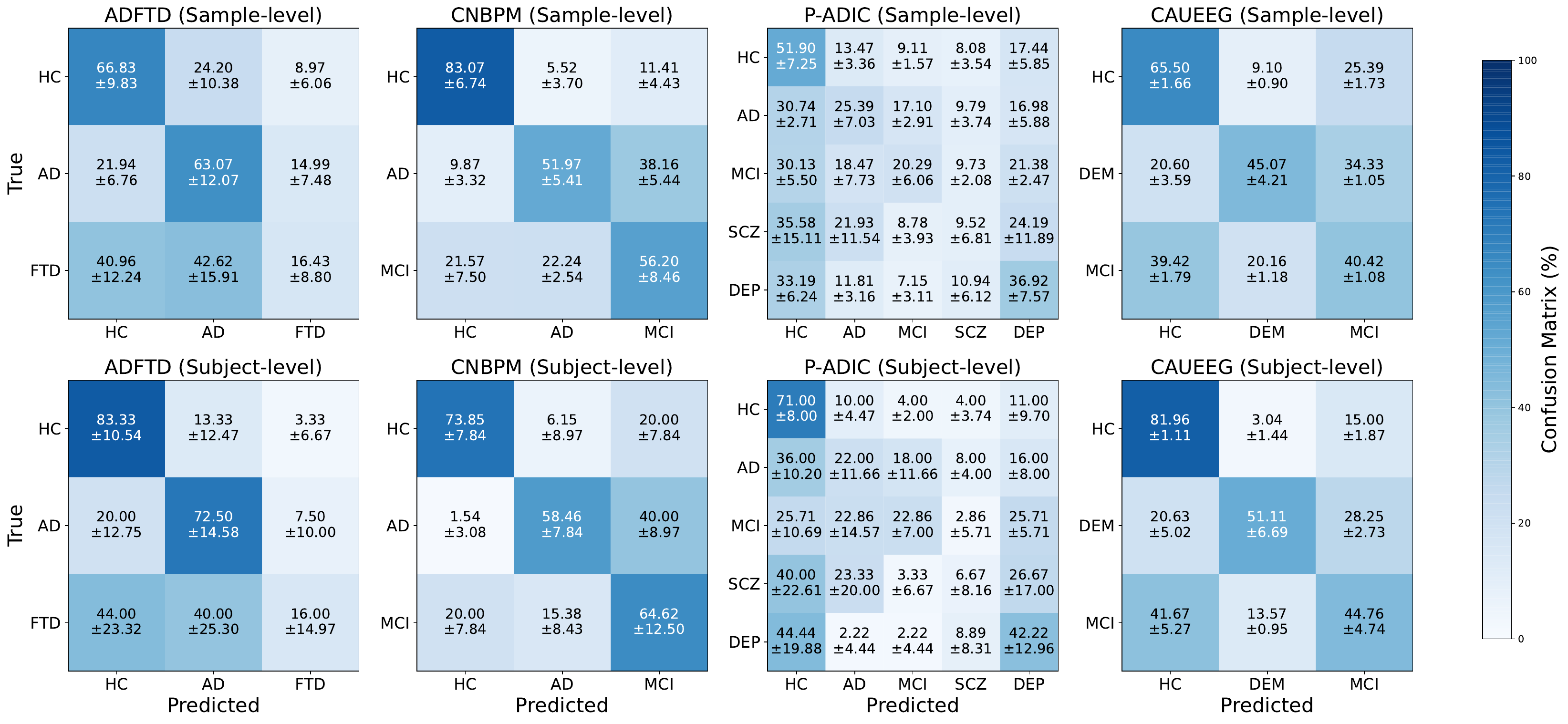}
    \caption{\textbf{Confusion Matrix}. Sample- and subject-level confusion matrix for multi-class classification results on 4 datasets.
    }
    \label{fig:confusion_matrix}
\end{figure*}

\subsection{Training Settings}
\label{sub:training_setting}
We train the model for up to 200 epochs, with early stopping based on validation performance and a patience of 15 epochs, determined by the best sample-level F1 score. We use a batch size of 512. The optimizer is AdamW with a learning rate of 1e-4. A CosineAnnealingLR scheduler is used to dynamically adjust the learning rate during training. Model performance is reported using accuracy and F1 score at both the sample and subject levels, with subject-level predictions obtained through majority voting as described in Section~\ref{sub:eeg_disease_detection}. Additional metrics, including precision, sensitivity, specificity, AUROC, and AUPRC, are also computed and available through our publicly released evaluation code but are omitted from the tables for conciseness. We adopt a Monte Carlo cross-validation strategy~\cite{wang2026lead} with a subject-independent train/validation/test split ratio of 6:2:2 for each dataset. We repeat each experiment with five different random seeds (41-45) and report the mean and standard deviation of the evaluation metrics. All experiments are conducted on a cluster with four NVIDIA A5000 GPUs using Python 3.8 and PyTorch 2.0.0+cu118. Further implementation details are provided in the scripts in our GitHub repository.

\section{Results and Analysis}
\label{sec:experimental_results}

\begin{table*}[!t]
    \centering
    \caption{\textbf{AD/Dementia vs HC Results.} The DEM vs HC in CAUEEG and AD vs HC in the other 3 datasets.
    }
    \label{tab:ad_vs_hc_tasks}
    \resizebox{\textwidth}{!}{
    \begin{tabular}{@{}ll|cc|cc|cc|cc@{}}
    \toprule

    \multicolumn{2}{l|}{\textbf{Datasets}}
    & \multicolumn{2}{c|}{\makecell{\textbf{ADFTD} \\ \textit{(142,966 Samples)}  \\ \textit{(65 Subjects, 2 Classes)} }}
    & \multicolumn{2}{c|}{\makecell{\textbf{CNBPM}  \\ \textit{(92,619 Samples)}  \\ \textit{(126 Subjects, 2 Classes)} }}
    & \multicolumn{2}{c|}{\makecell{\textbf{P-ADIC} \\ \textit{(326,898  Samples)}  \\ \textit{(145 Subjects, 2 Classes)} }}
    & \multicolumn{2}{c}{\makecell{\textbf{CAUEEG} \\ \textit{(1,074,469 Samples)}  \\ \textit{(770 Subjects, 2 Classes)} }}
    \\ \midrule

    \multicolumn{2}{l|}{\diagbox{\textbf{Methods}}{\textbf{Metrics}}} & \textbf{Accuracy} & \textbf{F1 Score} & \textbf{Accuracy} & \textbf{F1 Score} & \textbf{Accuracy} & \textbf{F1 Score} & \textbf{Accuracy} & \textbf{F1 Score} \\ \midrule

    \multicolumn{10}{c}{\textbf{Sample-Level Classification}}  \\
    \midrule

    \multicolumn{2}{l|}{\textbf{ManualFeature}}  & 61.88\std{4.44} & 61.61\std{4.64} & 68.78\std{4.65} & 63.78\std{5.58} & 62.14\std{1.35} & 57.06\std{2.26} & 70.32\std{1.28} & 69.91\std{1.40}  \\
    \multicolumn{2}{l|}{\textbf{EEGNet}}  & 63.05\std{5.44} & 62.64\std{5.72} & 71.38\std{7.36} & 67.83\std{7.09} & \textbf{67.30\std{3.86}} & 60.74\std{4.47} & 68.97\std{0.99} & 68.45\std{1.15}  \\
    \multicolumn{2}{l|}{\textbf{TST}}  & 65.58\std{4.11} & 65.25\std{4.01} & 85.32\std{5.34} & 83.24\std{6.24} & 62.89\std{3.09} & 60.03\std{2.23} & 73.86\std{1.60} & 73.24\std{1.89}  \\
    \multicolumn{2}{l|}{\textbf{EEGInception}}  & \textbf{73.66\std{3.90}} & \textbf{73.51\std{4.00}} & 86.08\std{5.12} & 84.32\std{5.10} & 65.31\std{2.10} & 59.44\std{2.02} & 75.22\std{1.51} & 74.75\std{1.94}  \\
    \multicolumn{2}{l|}{\textbf{EEGConformer}}  & 72.12\std{3.27} & 71.77\std{3.11} & 86.86\std{5.87} & 85.38\std{6.17} & 63.78\std{3.39} & 59.26\std{2.58} & \textbf{76.73\std{1.27}} & \textbf{76.09\std{1.61}}  \\
    \multicolumn{2}{l|}{\textbf{BIOT}}  & 69.37\std{5.10} & 68.99\std{5.00} & 82.30\std{7.38} & 79.38\std{7.75} & 63.39\std{2.74} & 57.96\std{2.66} & 73.75\std{1.11} & 73.03\std{1.24}  \\
    \multicolumn{2}{l|}{\textbf{MedGNN}}  & 70.39\std{3.09} & 70.20\std{2.91} & 86.78\std{4.52} & 84.57\std{5.73} & 64.09\std{2.62} & 59.10\std{1.66} & 75.70\std{1.60} & 74.97\std{1.97}  \\
    \multicolumn{2}{l|}{\textbf{Medformer}}  & 69.98\std{2.93} & 69.76\std{2.82} & 85.42\std{6.18} & 83.63\std{6.98} & 64.94\std{2.49} & 61.31\std{1.61} & 74.34\std{1.43} & 73.59\std{1.86}  \\
    \multicolumn{2}{l|}{\textbf{ModernTCN}}  & 67.49\std{6.24} & 66.86\std{6.59} & 78.79\std{4.51} & 76.32\std{4.80} & 62.93\std{6.15} & 55.57\std{5.83} & 74.31\std{1.26} & 73.99\std{1.27}  \\
    \multicolumn{2}{l|}{\textbf{ADformer (Ours)}}  & 72.59\std{4.41} & 72.52\std{4.38} & \textbf{88.05\std{3.76}} & \textbf{86.21\std{4.56}} & 65.34\std{3.57} & \textbf{61.65\std{2.25}} & 76.21\std{1.62} & 75.62\std{1.96}  \\

    \midrule

    \multicolumn{10}{c}{\textbf{Subject-Level Detection}}  \\
    \midrule

    \multicolumn{2}{l|}{\textbf{ManualFeature}}  & 72.86\std{8.33} & 71.90\std{8.82} & 73.08\std{11.67} & 69.97\std{15.96} & 68.00\std{3.40} & 59.18\std{6.92} & 80.39\std{1.76} & 79.39\std{1.84}  \\
    \multicolumn{2}{l|}{\textbf{EEGNet}}  & 70.00\std{12.29} & 68.78\std{12.89} & 79.23\std{6.71} & 78.34\std{7.84} & 72.00\std{4.99} & 61.83\std{8.38} & 77.03\std{1.61} & 75.87\std{2.14}  \\
    \multicolumn{2}{l|}{\textbf{TST}}  & 68.57\std{10.69} & 67.13\std{11.12} & 84.62\std{10.03} & 84.39\std{10.09} & 68.67\std{4.00} & 64.71\std{5.26} & 83.48\std{1.66} & 82.14\std{2.30}  \\
    \multicolumn{2}{l|}{\textbf{EEGInception}}  & 81.43\std{9.69} & 81.27\std{9.80} & 86.92\std{5.76} & 86.63\std{6.21} & 71.33\std{4.00} & 64.43\std{3.70} & 83.74\std{1.70} & 82.72\std{2.28}  \\
    \multicolumn{2}{l|}{\textbf{EEGConformer}}  & 80.00\std{5.35} & 79.10\std{5.38} & 87.69\std{10.15} & 87.64\std{10.21} & 68.00\std{4.52} & 60.00\std{7.40} & 85.29\std{1.93} & 84.22\std{2.37}  \\
    \multicolumn{2}{l|}{\textbf{BIOT}}  & 77.14\std{8.33} & 76.52\std{8.62} & 83.08\std{5.76} & 82.97\std{5.87} & 71.33\std{4.52} & 63.88\std{8.68} & 83.35\std{1.80} & 81.99\std{2.01}  \\
    \multicolumn{2}{l|}{\textbf{MedGNN}}  & 80.00\std{5.35} & 79.58\std{5.30} & 86.92\std{6.25} & 86.89\std{6.24} & 67.33\std{4.90} & 61.28\std{3.41} & 84.00\std{2.43} & 82.69\std{3.04}  \\
    \multicolumn{2}{l|}{\textbf{Medformer}}  & 82.86\std{5.71} & 82.72\std{5.64} & 85.38\std{8.21} & 85.08\std{8.40} & 71.33\std{1.63} & 66.95\std{2.91} & 83.10\std{1.70} & 81.60\std{2.31}  \\
    \multicolumn{2}{l|}{\textbf{ModernTCN}}  & 71.43\std{11.07} & 69.58\std{11.49} & 85.38\std{4.49} & 85.35\std{4.50} & 66.00\std{8.27} & 55.90\std{11.29} & 83.48\std{1.33} & 82.67\std{1.47}  \\
    \multicolumn{2}{l|}{\textbf{ADformer (Ours)}}  & \textbf{87.14\std{5.35}} & \textbf{87.11\std{5.36}} & \textbf{90.00\std{6.25}} & \textbf{89.83\std{6.48}} & \textbf{72.00\std{5.81}} & \textbf{67.99\std{6.57}} & \textbf{85.55\std{2.52}} & \textbf{84.42\std{3.06}}  \\

    \bottomrule
    \end{tabular}
    }
\end{table*}

\subsection{Multi-Class Classification}
\label{sub:multi_class_results}

We begin by performing a multi-class classification task to evaluate our model's ability to detect AD as well as other brain disorders, such as FTD, MCI, Schizophrenia (SCZ), and Depression (DEP). Table~\ref{tab:processed_data} summarizes the disorder classes for each dataset. It is worth noting that for the CAUEEG dataset, AD subjects are included within the broader Dementia (DEM) category, which also encompasses other dementia types such as vascular dementia. We retain the original class labels provided by the dataset. In real-world clinical scenarios, detecting brain disorders is far more complex than a standard multi-class classification task. One challenge is that a single patient may have comorbid neurological conditions, making the task inherently multi-label. In such cases, each disease class may require a distinct projection head for prediction. Another complication arises from overlapping disorder categories. For example, some MCI patients eventually progress to AD (referred to as progressive MCI or PMCI), while others remain stable (SMCI) or revert to a healthy condition~\cite{vecchio2018sustainable,ge2025eeg}. Importantly, PMCI subjects may already exhibit AD-related features at the time of data collection, yet without longitudinal follow-up studies, their future diagnosis remains unknown. In this paper, we simplify the task to a multi-class classification for model evaluation. However, this simplification does not reflect the complexity of real-world scenarios and should be used with caution in practical deployments.

Table~\ref{tab:multi_class_tasks} presents the experimental results, and Figure~\ref{fig:confusion_matrix} illustrates the confusion matrix. Our method achieves the best performance in 12 out of 16 reported metrics across the four datasets. In particular, for subject-level metrics, including accuracy and F1 score, our method consistently achieves the top results on all datasets, demonstrating its potential utility in real-world detection settings. On ADFTD, our method only correctly identifies 16\% of FTD cases at the subject level and classifies the remaining cases as HC (44\%) or AD (40\%), suggesting that the model struggles to identify EEG patterns that distinguish FTD from both groups. The results from CNBPM and CAUEEG support our earlier hypothesis that distinguishing MCI from AD/Dementia can be challenging, likely due to unobserved overlap at the time of data acquisition. The relatively poor performance on P-ADIC suggests that distinguishing between heterogeneous brain disorders such as SCZ and DEP, when mixed with dementia-related conditions like AD, HC, and MCI, remains difficult. In clinical practice, clinicians may benefit from applying preliminary diagnostic filtering to exclude non-dementia-related subjects when the target is AD detection. Interestingly, we observe that the best sample-level performance does not always mean the best subject-level results after majority voting. For example, EEGInception achieves the highest sample-level F1 score on the ADFTD dataset, outperforming our method by 1.24\%; however, our method surpasses EEGInception at the subject level by 3.01\% in F1 score. This discrepancy highlights a potential issue: post-processing methods such as majority voting are not incorporated into model training, so sample-level results may not align with subject-level performance. This suggests future work on subject-level training strategies to better bridge the gap between sample-level training objectives and subject-level evaluation in clinical applications.

\subsection{AD vs HC Classification}
\label{sub:ad_vs_hc_results}

We then perform a binary classification task to evaluate the model’s ability to distinguish between AD and HC subjects. AD is the most prevalent neurodegenerative disease among the elderly and is the primary cause of dementia. As noted previously, in the CAUEEG dataset, AD subjects are grouped under the broader Dementia category, which includes other forms of dementia such as VD. In contrast, the other three datasets used in this study provide a distinct AD class, allowing for more precise evaluation of AD-specific detection.

Table~\ref{tab:ad_vs_hc_tasks} reports the results. Our method achieves the best performance on 11 of 16 metrics and the best subject-level F1 scores across the four datasets. In particular, subject-level F1 scores reach about 90\% for ADFTD and CNBPM and exceed 80\% for CAUEEG, indicating that EEG-based methods can effectively capture discriminative patterns between AD and HC subjects. The relatively lower performance on P-ADIC may reflect data quality issues, such as artifact contamination or inconsistent data collection. Additionally, we again observe that the highest sample-level performance does not necessarily yield the best subject-level results, consistent with our findings from the multi-class classification experiments. This reinforces the importance of evaluating both levels and calls for further research into subject-level alignment and training strategies to improve consistency between sample-level learning and subject-level outcomes.

\subsection{HC vs AD vs VD Classification}
\label{sub:hc_ad_vd_results}

As discussed previously, the CAUEEG dataset defines a broad dementia category comprising AD, VD, and a small number of FTD and Parkinson’s disease dementia cases. To further evaluate the model’s ability to distinguish AD from other dementia types, we refine the dementia class by separating AD and VD subjects and excluding other subtypes because they have very few subjects. We then conduct a three-class classification task of HC vs AD vs VD.

\begin{table}[!t]
    \centering
    \caption{\textbf{HC vs AD vs VD Results.} The HC vs AD vs VD classification results of the CAUEEG dataset.}
    \label{tab:hc_ad_vd_tasks}
    \resizebox{\columnwidth}{!}{
    \begin{tabular}{@{}ll|cc|cc@{}}
    \toprule

    \multicolumn{2}{l|}{\textbf{Datasets}}  
    & \multicolumn{4}{c}{\makecell{\textbf{CAUEEG} \\ \textit{(1,071,111 Samples)}  \\ \textit{(769 Subjects, 3 Classes)} }}  \\
    \midrule

    \multicolumn{2}{l|}{} & \multicolumn{2}{c|}{\textbf{Sample-Level}}  & \multicolumn{2}{c}{\textbf{Subject-Level}}  \\  \midrule
    \multicolumn{2}{l|}{\diagbox{\textbf{Methods}}{\textbf{Metrics}}} & \textbf{Accuracy} & \textbf{F1 Score} & \textbf{Accuracy} & \textbf{F1 Score} \\ \midrule

    \multicolumn{2}{l|}{\textbf{ManualFeature}}  & 61.87\std{1.93} & 47.58\std{1.50}   & 72.08\std{2.05} & 47.23\std{1.84} \\
    \multicolumn{2}{l|}{\textbf{EEGNet}}  & 60.42\std{2.20} & 45.18\std{5.08}   & 68.57\std{3.35} & 48.95\std{7.85} \\
    \multicolumn{2}{l|}{\textbf{TST}}  & 64.98\std{1.00} & 54.55\std{1.73}   & 74.94\std{2.27} & 61.11\std{3.94} \\
    \multicolumn{2}{l|}{\textbf{EEGInception}}  & 66.39\std{1.51} & 56.66\std{1.91}   & 76.75\std{2.11} & 66.00\std{2.80} \\
    \multicolumn{2}{l|}{\textbf{EEGConformer}}  & \textbf{67.92\std{0.48}} & 57.22\std{1.46}   & 76.23\std{2.92} & 65.38\std{4.70} \\
    \multicolumn{2}{l|}{\textbf{BIOT}}  & 65.81\std{1.55} & 56.72\std{1.23}   & \textbf{77.66\std{1.77}} & 66.39\std{2.37} \\
    \multicolumn{2}{l|}{\textbf{MedGNN}}  & 67.45\std{0.75} & 56.42\std{1.89}   & 76.49\std{1.81} & 63.51\std{4.35} \\
    \multicolumn{2}{l|}{\textbf{Medformer}}  & 66.52\std{0.66} & 55.00\std{1.62}   & 75.58\std{1.46} & 60.33\std{5.05} \\
    \multicolumn{2}{l|}{\textbf{ModernTCN}}  & 65.30\std{1.18} & 55.17\std{2.24}   & 77.14\std{1.71} & 61.87\std{6.07} \\
    \multicolumn{2}{l|}{\textbf{ADformer (Ours)}}  & 67.60\std{0.21} & \textbf{57.27\std{0.78}}   & 77.27\std{1.30} & \textbf{66.76\std{2.37}} \\
            
    \bottomrule
    \end{tabular}
    }
\end{table}

Table~\ref{tab:hc_ad_vd_tasks} summarizes the results. Our method achieves the highest F1 scores at both the sample and subject levels. Notably, the subject-level F1 for HC vs AD vs VD (66.76\%) exceeds that for HC vs MCI vs DEM (59.20\%), indicating that differentiating dementia subtypes may be easier than distinguishing MCI from either HC or dementia. 
As discussed in Section~\ref{sub:multi_class_results}, this performance gap likely stems from overlapping distributions between MCI and dementia categories.

\subsection{Sample Length \& Overlapping Study}
\label{sub:length_overlap_study}

\begin{table}[!t]
    \centering
    \caption{\textbf{Sample Length Study.} We evaluate how different lengths $T$ impact the performance on the AD/Dementia vs HC task.}
    \label{tab:sample_length_study}
    \resizebox{\columnwidth}{!}{
    \begin{tabular}{@{}ll|cc|cc@{}}
    \toprule
    \multicolumn{2}{l|}{\textbf{Datasets}}
    & \multicolumn{2}{c|}{\makecell{\textbf{ADFTD} \\ \textit{(142,966 Samples)} \\ \textit{(65 Subjects, 2 Classes)}}}
    & \multicolumn{2}{c}{\makecell{\textbf{CAUEEG} \\ \textit{(1,074,469 Samples)} \\ \textit{(770 Subjects, 2 Classes)}}} \\
    \midrule
    \multicolumn{2}{l|}{\diagbox{\textbf{Lengths}}{\textbf{Metrics}}}
    & \textbf{Accuracy} & \textbf{F1 Score} & \textbf{Accuracy} & \textbf{F1 Score} \\
    \midrule
    \multicolumn{6}{c}{\textbf{Sample-Level Classification}} \\
    \midrule
    \multicolumn{2}{l|}{\textbf{128 (1s)}}  & 71.61\std{4.57} & \textbf{71.50\std{4.59}} & 75.70\std{1.43} & 74.98\std{1.74} \\
    \multicolumn{2}{l|}{\textbf{256 (2s)}}  & 70.77\std{5.99} & 70.66\std{6.06} & \textbf{76.04\std{1.56}} & \textbf{75.34\std{2.00}} \\
    \multicolumn{2}{l|}{\textbf{512 (4s)}}  & \textbf{71.64\std{5.43}} & 71.46\std{5.51} & 75.64\std{1.80} & 74.77\std{2.12} \\
    \multicolumn{2}{l|}{\textbf{1024 (8s)}} & 68.61\std{5.07} & 68.27\std{4.97} & 75.02\std{1.91} & 74.20\std{2.20} \\
    \midrule
    \multicolumn{6}{c}{\textbf{Subject-Level Detection}} \\
    \midrule
    \multicolumn{2}{l|}{\textbf{128 (1s)}}  & 80.00\std{5.35} & 79.84\std{5.55} & \textbf{84.52\std{2.34}} & \textbf{83.31\std{2.65}} \\
    \multicolumn{2}{l|}{\textbf{256 (2s)}}  & \textbf{81.43\std{8.57}} & \textbf{81.33\std{8.62}} & 83.87\std{1.87} & 82.51\std{2.48} \\
    \multicolumn{2}{l|}{\textbf{512 (4s)}}  & 77.14\std{5.35} & 76.23\std{5.90} & 82.19\std{1.94} & 80.68\std{2.19} \\
    \multicolumn{2}{l|}{\textbf{1024 (8s)}} & 78.57\std{11.95} & 78.04\std{12.16} & 80.90\std{1.33} & 79.24\std{1.62} \\
    \bottomrule
    \end{tabular}
    }
\end{table}

\begin{figure}[!t]
    \centering
    \includegraphics[width=1.0\linewidth]{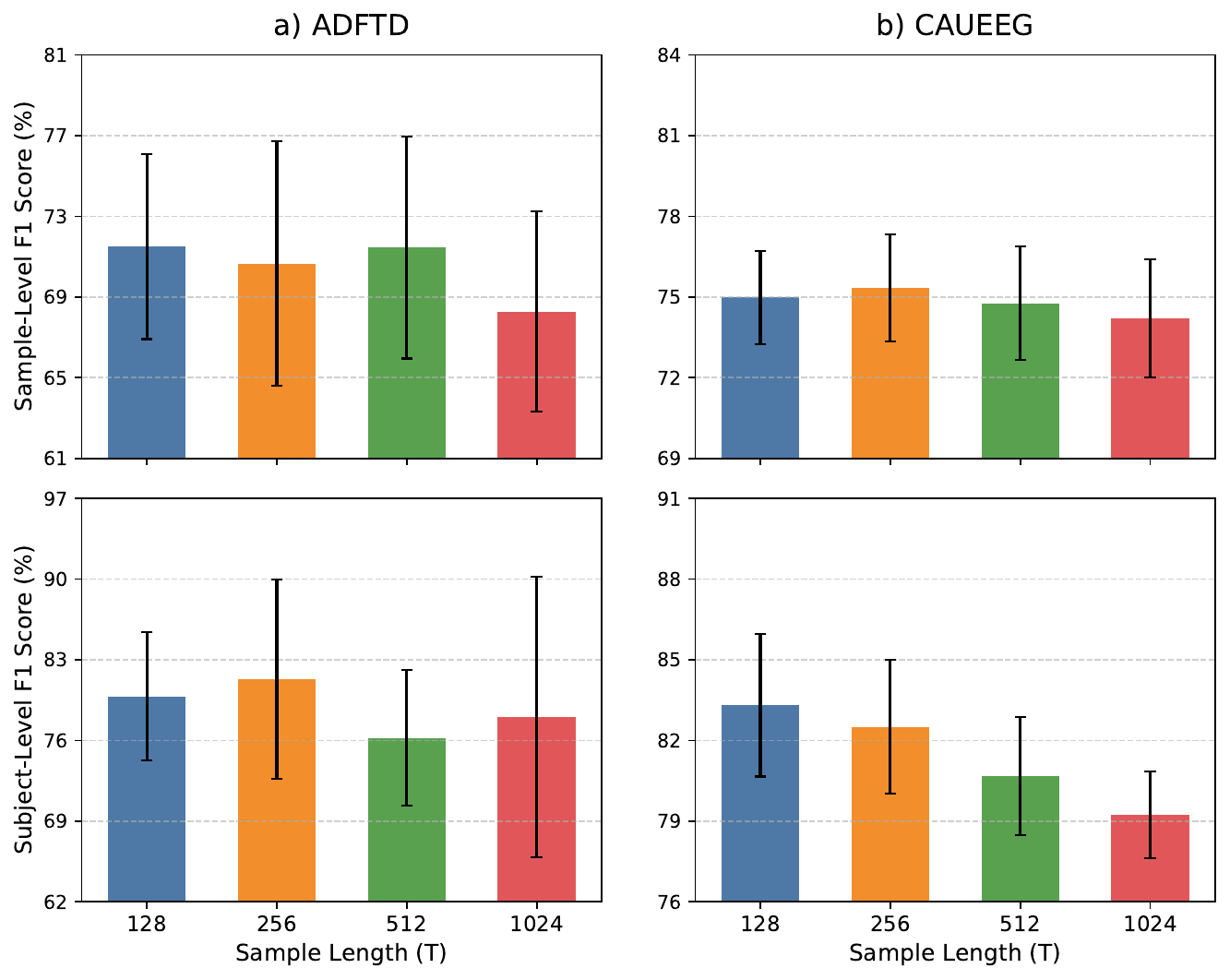}
    \caption{\textbf{Sample Lengths.} F1 score comparison for different sample lengths $T$.
    }
    \label{fig:sample_length_study}
\end{figure}

Sample length and overlap rate are two critical parameters in EEG data segmentation. Appropriate choices can preserve essential temporal and frequency features while increasing the amount of training data available for deep learning models. Since our ultimate goal is to determine whether a subject has AD using aggregated sample-level predictions, it is important to examine how varying sample lengths and overlapping rates affect subject-level performance in EEG-based AD detection. To this end, we evaluate the binary classification task of AD/Dementia vs. HC on two publicly available datasets, ADFTD and CAUEEG, which have shown relatively strong performance in this task. We evaluate 4 different sample lengths: 128(1s), 256(2s), 512(4s), and 1024(8s) timestamps. We then fix the sample length at 128 and vary the overlap rate at 0\%, 20\%, 50\%, and 80\%. For simplicity, we fix the patch length $L_i$ to 4 and the scaled channel number $F_j$ to 76, adopting a single-granularity setup across all experiments.

\begin{table}[!t]
    \centering
    \caption{\textbf{Overlapping Rate Study.} We evaluate how different sample overlapping ratios impact the performance of ADFTD and CAUEEG on the AD/Dementia vs HC task.}
    \label{tab:sample_overlap_study}
    \resizebox{\columnwidth}{!}{
    \begin{tabular}{@{}ll|cc|cc@{}}
    \toprule
    \multicolumn{2}{l|}{\textbf{Datasets}}
    & \multicolumn{2}{c|}{\makecell{\textbf{ADFTD} \\ \textit{(142,966 Samples)} \\ \textit{(65 Subjects, 2 Classes)}}}
    & \multicolumn{2}{c}{\makecell{\textbf{CAUEEG} \\ \textit{(1,074,469 Samples)} \\ \textit{(770 Subjects, 2 Classes)}}} \\
    \midrule
    \multicolumn{2}{l|}{\diagbox{\textbf{Overlap}}{\textbf{Metrics}}} & \textbf{Accuracy} & \textbf{F1 Score} & \textbf{Accuracy} & \textbf{F1 Score} \\
    \midrule
    \multicolumn{6}{c}{\textbf{Sample-Level Classification}} \\
    \midrule
    \multicolumn{2}{l|}{\textbf{0\%}} & 70.93\std{4.42} & 70.73\std{4.49} & 75.00\std{1.78} & 74.26\std{2.08} \\
    \multicolumn{2}{l|}{\textbf{20\%}} & 70.70\std{6.41} & 70.57\std{6.43} & 75.09\std{1.90} & 74.24\std{2.34} \\
    \multicolumn{2}{l|}{\textbf{50\%}} & 71.61\std{4.57} & 71.50\std{4.59} & \textbf{75.70\std{1.43}} & \textbf{74.98\std{1.74}} \\
    \multicolumn{2}{l|}{\textbf{80\%}} & \textbf{72.59\std{2.85}} & \textbf{72.52\std{2.86}} & 75.50\std{1.39} & 74.83\std{1.67} \\
    \midrule
    \multicolumn{6}{c}{\textbf{Subject-Level Detection}} \\
    \midrule
    \multicolumn{2}{l|}{\textbf{0\%}} & 80.00\std{8.33} & 79.67\std{8.53} & 83.61\std{2.96} & 82.21\std{3.51} \\
    \multicolumn{2}{l|}{\textbf{20\%}} & 77.14\std{15.25} & 76.91\std{15.35} & 83.61\std{2.96} & 82.02\std{3.63} \\
    \multicolumn{2}{l|}{\textbf{50\%}} & 80.00\std{5.35} & 79.84\std{5.55} & 84.52\std{2.34} & 83.31\std{2.65} \\
    \multicolumn{2}{l|}{\textbf{80\%}} & \textbf{81.43\std{5.71}} & \textbf{81.23\std{5.89}} & \textbf{84.65\std{1.75}} & \textbf{83.49\std{2.17}} \\
    \bottomrule
    \end{tabular}
    }
\end{table}

\begin{figure}[!t]
    \centering
    \includegraphics[width=1.0\linewidth]{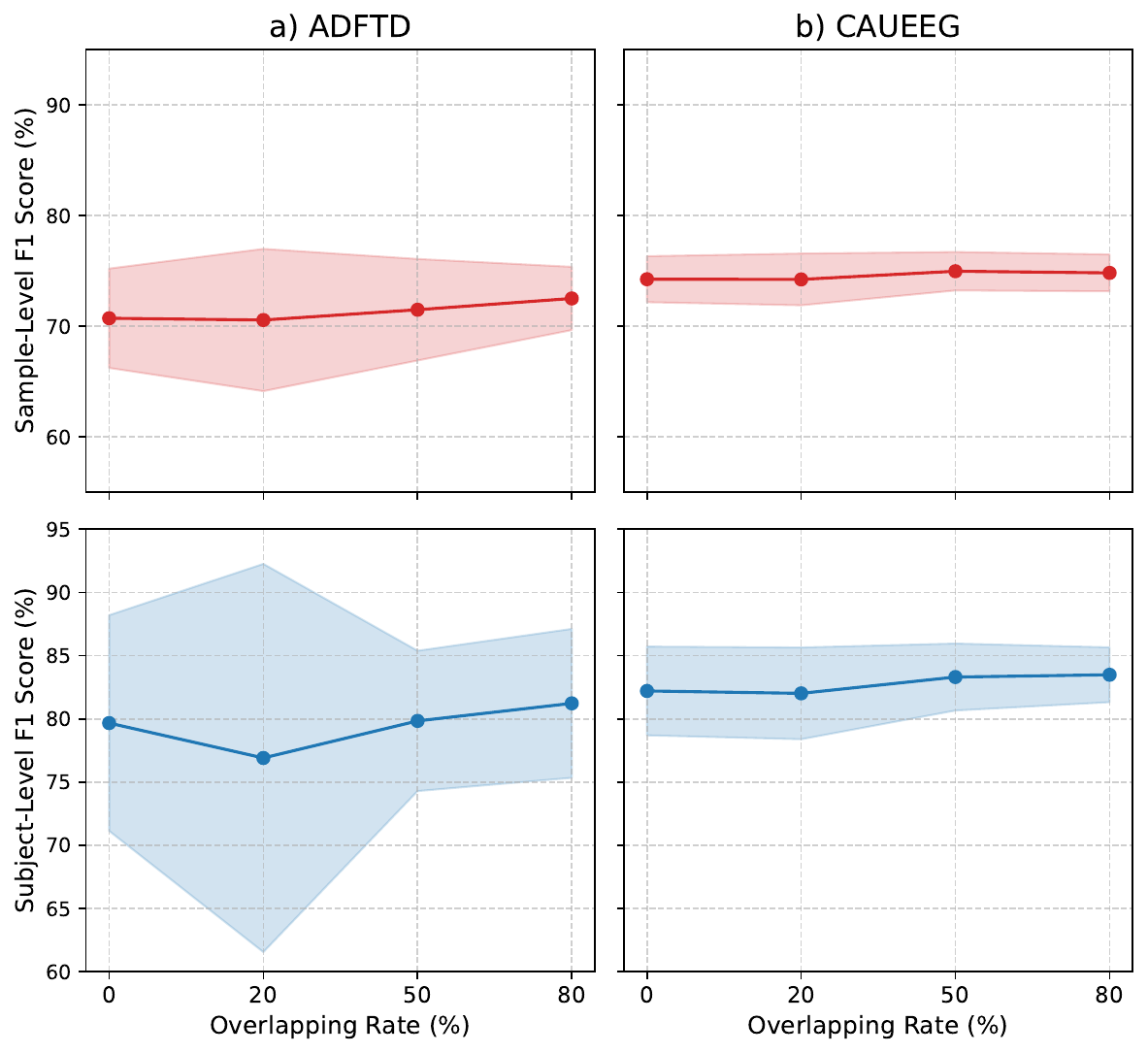}
    \caption{\textbf{Overlap Rates.} F1 score trends across different overlap rates.
    }
    \label{fig:overlap_rate_study}
\end{figure}

The results are summarized in Table~\ref{tab:sample_length_study} and Table~\ref{tab:sample_overlap_study}. Figure~\ref{fig:sample_length_study} and Figure~\ref{fig:overlap_rate_study} further illustrate performance comparisons and trends under varying sample lengths and overlap rates. For sample length, the best performance, in both sample-level and subject-level F1 scores, occurs at 128 and 256 timestamps on both datasets. \textit{This suggests that arbitrarily increasing sample length does not necessarily improve overall performance}, likely because it reduces the number of training samples, which can hurt learning. For the overlap rate, we observe that 80\% overlap yields the 6 of 8 best results for sample-level and subject-level metrics across datasets. The 80\% overlap setting also has the smallest standard deviation across five runs in all four comparisons. Overall, \textit{both 50\% and 80\% overlap yield higher mean F1 scores than 0\% and 20\% overlap at both evaluation levels on both datasets.}

\begin{table*}[!t]
    \centering
    \scriptsize
    \caption{\textbf{Ablation Study.} Effectiveness of each module of our method on the AD/Dementia vs HC task.
    }
    \label{tab:module_study}
    \resizebox{\textwidth}{!}{
    \begin{tabular}{@{}ll|cc|cc|cc|cc@{}}
    \toprule

    \multicolumn{2}{l|}{\textbf{Datasets}}
    & \multicolumn{2}{c|}{\makecell{\textbf{ADFTD} \\ \textit{(142,966 Samples)}  \\ \textit{(65 Subjects, 2 Classes)} }}
    & \multicolumn{2}{c|}{\makecell{\textbf{CNBPM}  \\ \textit{(92,619 Samples)}  \\ \textit{(126 Subjects, 2 Classes)} }}
    & \multicolumn{2}{c|}{\makecell{\textbf{P-ADIC} \\ \textit{(326,898  Samples)}  \\ \textit{(145 Subjects, 2 Classes)} }}
    & \multicolumn{2}{c}{\makecell{\textbf{CAUEEG} \\ \textit{(1,074,469 Samples)}  \\ \textit{(770 Subjects, 2 Classes)} }}
    \\ \midrule

    \multicolumn{2}{l|}{\diagbox{\textbf{Methods}}{\textbf{Metrics}}} & \textbf{Accuracy} & \textbf{F1 Score} & \textbf{Accuracy} & \textbf{F1 Score} & \textbf{Accuracy} & \textbf{F1 Score} & \textbf{Accuracy} & \textbf{F1 Score} \\ \midrule
    \multicolumn{10}{c}{\textbf{Sample-Level Classification}} \\
    \midrule
    \multicolumn{2}{l|}{\textbf{No Inter-Attention}} & \textbf{72.89\std{4.62}} & \textbf{72.79\std{4.58}} & 86.36\std{4.67} & 84.57\std{5.40} & \textbf{65.80\std{3.87}} & \textbf{61.80\std{1.90}} & 76.13\std{1.60} & 75.56\std{1.88} \\
    \multicolumn{2}{l|}{\textbf{No Temporal Branch}} & 71.03\std{3.61} & 70.59\std{3.43} & 86.43\std{3.46} & 84.19\std{3.97} & 65.13\std{2.00} & 57.55\std{5.36} & 75.56\std{1.28} & 74.84\std{1.61} \\
    \multicolumn{2}{l|}{\textbf{No Spatial Branch}} & 70.01\std{4.18} & 69.94\std{4.20} & 87.50\std{4.62} & 85.79\std{5.31} & 64.82\std{4.61} & 61.75\std{3.10} & 75.39\std{1.40} & 74.66\std{1.75} \\
    \multicolumn{2}{l|}{\textbf{All Module Utilized}} & 72.59\std{4.41} & 72.52\std{4.38} & \textbf{88.05\std{3.76}} & \textbf{86.21\std{4.56}} & 65.34\std{3.57} & 61.65\std{2.25} & \textbf{76.21\std{1.62}} & \textbf{75.62\std{1.96}} \\

    \midrule
    \multicolumn{10}{c}{\textbf{Subject-Level Detection}} \\
    \midrule
    \multicolumn{2}{l|}{\textbf{No Inter-Attention}} & 81.43\std{13.25} & 81.03\std{13.61} & 86.15\std{4.62} & 85.98\std{4.78} & \textbf{73.33\std{4.22}} & \textbf{68.25\std{5.17}} & 85.16\std{1.87} & 84.05\std{2.27} \\
    \multicolumn{2}{l|}{\textbf{No Temporal Branch}} & 75.71\std{5.71} & 75.02\std{5.35} & 86.15\std{5.22} & 86.11\std{5.22} & 67.33\std{5.73} & 54.98\std{12.97} & 83.87\std{2.86} & 82.57\std{3.33} \\
    \multicolumn{2}{l|}{\textbf{No Spatial Branch}} & 80.00\std{8.33} & 79.73\std{8.57} & 89.23\std{6.15} & 88.96\std{6.67} & 71.33\std{7.48} & 67.13\std{7.23} & 83.87\std{2.74} & 82.50\std{3.19} \\
    \multicolumn{2}{l|}{\textbf{All Module Utilized}} & \textbf{87.14\std{5.35}} & \textbf{87.11\std{5.36}} & \textbf{90.00\std{6.25}} & \textbf{89.83\std{6.48}} & 72.00\std{5.81} & 67.99\std{6.57} & \textbf{85.55\std{2.52}} & \textbf{84.42\std{3.06}} \\

    \bottomrule
    \end{tabular}
    }
\end{table*}

\begin{figure*}[!t]
    \centering
    \includegraphics[width=1.0\linewidth]{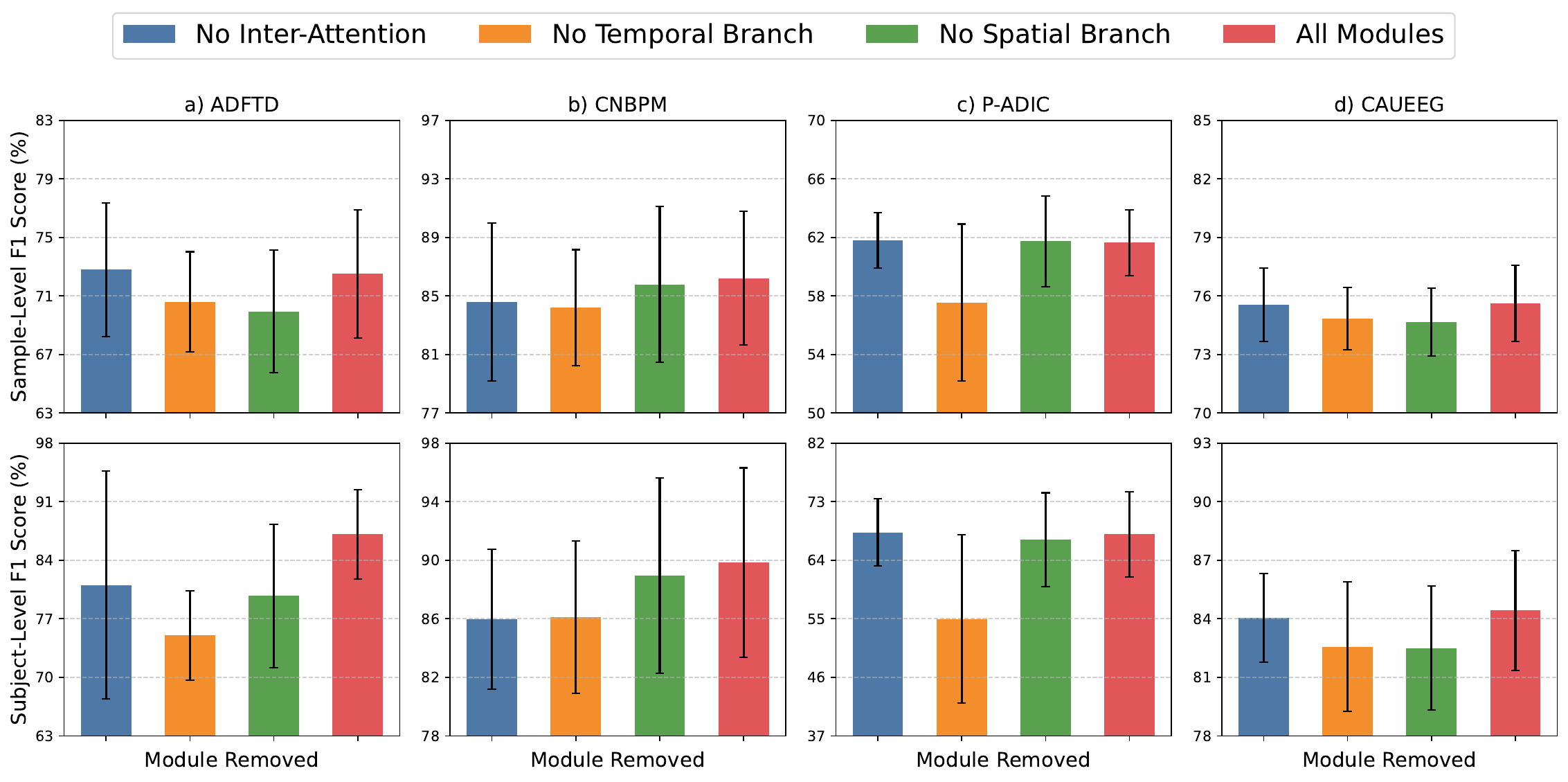}
    \caption{\textbf{Ablation Study.} F1 score of ablation studies across different model variants by removing modules on the AD/Dementia vs HC task.
    }
    \label{fig:module_ablation_study}
\end{figure*}

\subsection{Ablation Study}
\label{sub:ablation_study}

To evaluate each component in our proposed method, we conduct an ablation study by systematically removing individual modules and assessing the resulting performance changes. Specifically, we evaluate the impact of removing the inter-granularity self-attention, the temporal branch, and the spatial branch. The performance of the complete model (with all modules included) is also reported for comparison. All modules use the same parameter choices as in Table~\ref{tab:ad_vs_hc_tasks}.

Table~\ref{tab:module_study} summarizes the results, and Figure~\ref{fig:module_ablation_study} visually compares the different model variants. For 3 out of the 4 datasets, the full model achieves the highest subject-level F1 score, demonstrating the complementary contributions of the proposed components. Across all datasets, both the temporal and spatial branches significantly improve performance; removing either branch causes a clear drop, highlighting the importance of modeling both temporal dynamics and spatial channel relationships in EEG data. Notably, the inter-granularity self-attention module does not consistently improve performance. For instance, on the P-ADIC dataset, removing it yields comparable or even slightly better results. This may be because this dataset has few granularity levels, making it difficult for the model to learn meaningful inter-granularity relationships. Overall, the ablation study confirms that each module design contributes positively to performance in most scenarios, validating the architecture choices.

\section{Discussion}
\label{sec:discussion}

\textbf{Performance Discussion.} To simulate real-world clinical scenarios, we employ a subject-independent cross-validation protocol and apply majority voting to aggregate sample-level predictions into subject-level detections. Our findings show that distinguishing dementia-related disorders from other brain disorders, such as schizophrenia and depression, remains challenging, indicating the potential need for pre-screening by clinicians to filter out unrelated diagnoses. In contrast, binary classification between AD and HC yields strong results, with subject-level F1 scores approaching 90\% across several datasets. Additionally, we explore how segmentation parameters, specifically sample length and overlapping rate, affect subject-level performance. Our results suggest that longer sample lengths do not necessarily yield better performance and may even degrade it. In contrast, applying sample overlap during segmentation, regardless of the specific overlap rate, consistently improves subject-level AD detection performance.

\textbf{Modality Comparison.} Established clinical diagnostic approaches, including MRI, PET, CSF analysis, and increasingly validated blood-based biomarkers, remain essential to AD diagnosis, with biomarker-based methods providing more accurate and disease-specific evidence of AD pathology than EEG-based approaches. Our goal is not to replace these established procedures but to explore EEG as a convenient, non-invasive, portable, and relatively low-cost detection tool. A similar example exists in blood pressure monitoring, where validated cuff-based measurements remain the clinical standard, while emerging wearable devices may offer more convenient long-term monitoring once their reliability is established. Likewise, EEG-based methods could potentially support accessible preliminary screening and longitudinal monitoring, including in home-based settings with wearable devices.

\textbf{Future Directions.} We identify four promising directions for future research: 1) \textit{Adaptive Granularity Selection}. In this work, we manually tune granularity levels for patch length and scaled channel number, which can be time-consuming and suboptimal across datasets. Any auto-adaptive granularity-selection mechanism may improve both flexibility and performance. 2) \textit{Subject-Level Regularization}. Our results reveal that the best sample-level performance does not always imply the best subject-level performance after majority voting. This discrepancy suggests the need for subject-aware training strategies. Future work could explore subject-level regularization methods that backpropagate subject-level prediction errors to guide model optimization. 3) \textit{Large-scale pre-training}. EEG foundation models using a Transformer backbone have recently emerged as a promising direction and show significant performance improvements on general EEG tasks. Given the cost of collecting EEG data for AD patients and the limited availability of EEG-AD datasets, it may be advisable to pre-train on subjects with other brain disorders and healthy controls to build generalizable EEG models, and then fine-tune on AD patients for detection. 4) \textit{Longitudinal EEG Data}. Since not all individuals with MCI eventually convert to AD, longitudinal EEG recordings are essential for reliably distinguishing PMCI from SMCI, thereby enabling more accurate early detection of AD.

\section{Conclusion}
\label{sec:conclusion}

This paper introduces \textbf{ADformer}, a novel end-to-end multi-granularity spatial-temporal transformer designed for EEG-based AD detection. We propose two innovative multi-granularity data embedding strategies that effectively capture both temporal and spatial features. In addition, we design a two-stage intra-inter granularity self-attention mechanism to extract local patterns within each granularity and global dependencies across granularities. To demonstrate the effectiveness of our method, 
we curate the world's largest EEG-based AD detection corpus by utilizing four large-scale datasets, including subjects with AD, MCI, FTD, healthy, and some other brain disorders. We conduct comprehensive evaluations on these four datasets, comprising a total of 1,713 subjects and 2,589,094 one-second samples. Experiments include multi-class classification among various brain disorders, binary classification between AD and HC subjects, and three-class classification among HC, AD, and VD to distinguish dementia. Compared with SOTA methods, our method achieves the best performance in 25 out of 36 metrics across the four datasets and three classification settings. Notably, ADformer achieves the best subject-level performance across all datasets and tasks, demonstrating strong real-world applicability for EEG-based AD detection.

\section*{References}
\bibliographystyle{IEEEtran}
\bibliography{refs}
\end{document}